\documentstyle[11pt,fleqn,leqno,epsfig,wrapfig]{article}
\begin{document}

\title{Very high-energy $\gamma$-ray observations of
the Crab nebula and other potential sources 
with the GRAAL experiment}
\author{        
         F.Arqueros$^1$, J.Ballestrin$^2$, M.Berenguel$^{3}$, D.M.Borque$^1$,
E.F.Camacho$^4$,\\  M.Diaz$^5$, H.-J.Gebauer$^{5}$, R.Enriquez$^{1}$,
R.Plaga$^{5}$\\  
 $^{1}$ Facultad de Ciencias Fisicas, Universidad Complutense,  
         E-28040 Madrid, Spain \\  
$^{2}$ CIEMAT-Departamento Energias Renovables, \\Plataforma Solar  
de Almeria, E-04080 Almeria, Spain\\    
$^{3}$ Departamento de Lenguajes y Computaci\'on, \\  
Universidad de Almeria, 04120 Almeria, Spain \\    
$^{4}$ Escuela Superior de Ingenieros, Universidad de Sevilla,  
E-41012 Sevilla, Spain \\               
$^{5}$ Max-Planck-Institut f\"ur Physik, 80805 M\"unchen, Germany} 
\maketitle
\begin{abstract}
\noindent
The ``Gamma Ray Astronomy at ALmeria'' (GRAAL) experiment
uses 63 heliostat-mirrors with a total mirror area
of  $\approx$ 2500 m$^2$
from the CESA-1 field at the ``Plataforma Solar de Almeria'' (PSA) to collect
Cherenkov light from air showers.
The detector is located in a central solar tower and 
detects photon-induced showers
with an energy threshold of 250 $\pm$ 110 GeV and an 
asymptotic effective 
detection area
of about 15000 m$^2$.
A comparison between the results
of detailed Monte-Carlo simulations and data is presented.
\\
Data sets taken
in the period September 1999 - September 2000 in the direction of
the Crab pulsar, the active galaxy 3C 454.3, the unidentified
$\gamma$-ray source 3EG 1835+35 and a ``pseudo source'' were analyzed
for high energy $\gamma$-ray emission.
Evidence for a $\gamma$-ray flux from the Crab pulsar with an
integral flux of 
2.2 $\pm$ 0.4 (stat) $^{+1.7}_{-1.3}$ (syst) $\times$ 
10$^{-9}$ cm$^{-2}$ sec$^{-1}$ above threshold
and a significance of 4.5 $\sigma$
in a total measuring time of 7 hours and 10 minutes on source
was found. No evidence for emission from the other sources was found.
\\
Some difficulties with the use of heliostat fields
for $\gamma$-ray astronomy are pointed out.
In particular
the effect of field-of-view restricted to the central part
of a detected air shower on the lateral distribution and timing
properties of Cherenkov light are discussed. Upon restriction
the spread of the timing front of proton induced showers
sharply decreases and the reconstructed direction becomes biased towards the
pointing direction.
This is shown to make efficient $\gamma$-hadron
separation difficult. 
\end{abstract}
\noindent
\section{Introduction - aims and plan of the paper}
\label{sec_intro}
Measuring atmospheric Cherenkov radiation 
is presently the most effective way to  detect
cosmic $\gamma$-rays with primary energies between about 100 GeV and 1 TeV
\cite{frank}. 
In order to reach low energy thresholds with techniques  
based on Cherenkov light, large mirror collection areas are needed.  
GRAAL is an experiment that employs  
the large mirror area of an existing tower
solar-power plant for this purpose.
\\
This paper briefly describes the GRAAL detector and reports results
about the detection of $\gamma$-rays from cosmic sources. In addition
some general lessons we learnt about the heliostat-field approach 
to $\gamma$-ray astronomy are reported.
In section \ref{detector} the GRAAL detector is described and compared
to other heliostat-field detectors for Cherenkov light.
Section \ref{evreco} describes the event reconstruction based mainly
on the arrival time of signals at the central detector. Section \ref{moncar}
treats the Monte Carlo simulation of the experiment. Section \ref{datasel}
explains how the data set used for the analysis of this paper
was chosen from the total set of all taken data. The data reduction 
procedures---and the fundamental problems besetting it---are explained 
in section \ref{datared} and the results are
presented in section \ref{results}. Finally some concluding remarks are offered
in section \ref{concl}. A more detailed report about these results
will be available in two theses\cite{davidthes,mariathes}.

\section{The GRAAL detector}
\label{detector}
\subsection{The CESA-1 heliostat field at the PSA}
\begin{figure}[ht]
\epsfig{
file=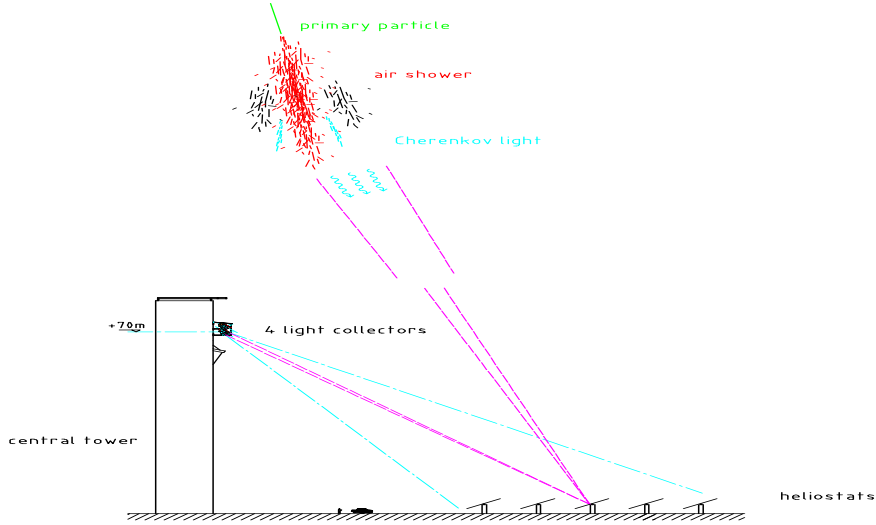,width=12cm,height=7cm,clip=,angle=0}
\caption{\it
Scheme of the experiment seen from the side, north is to the right.
The Cherenkov light of a schematic airshower 
(not to scale with respect to the field)
is concentrated by the heliostats of the CESA-1 field
to a focus at the central tower.
A dedicated platform mounted at the outside of the tower at the 70 m level  
houses four Winston  
cones which receive light from 13 - 18 heliostats in the field. 
The data-acquisition electronics is located inside the tower.
}
\label{tower}
\end{figure}

CESA-1 is a heliostat field comprising of 300 
steerable mirrors to the north
of a central tower located within the 
``Plataforma Solar de Almeria''(PSA) a solar thermal-energy 
research centre operated by the Spanish CIEMAT. 
The PSA is located in the desert of Tabernas 
(37$^{\circ}$.095 N, 2$^{\circ}$.360 W)
about 30 km from the city of Almeria and the sea, at the foothills of the 
Sierra-Nevada mountains (height a.s.l. of 505 m). 
The 63 heliostats used for GRAAL
have a mirror area of 39.7 m$^2$ each and consist of 12 rectangular ``facets''
(sub mirrors) with a spherical curvature
that are ``canted'' (adjusted relative to the overall frame)
to a roughly spherical overall heliostat shape.
The beam spread function of the heliostats has 
a RMS of about 0.25$^{\circ}$.
Each heliostat is individually steerable with stepping motors via
a central PC.
For the purpose of GRAAL
a control program was developed that allowed to perform the special
tracking needed for the use of the field for Cherenkov astronomy.
\\
The heliostat focus the Cherenkov light of air showers from
the direction of potential gamma-ray sources to 
software adjustable ``aiming points'' in the central tower 
(see fig.\ref{tower}). The so-called 
``convergent view''\cite{celesteprop}---the pointing of the heliostats
towards a point in the atmosphere corresponding
to an atmospheric depth of 230 g/cm$^2$ in the general direction of
the potential source of gamma rays---was always applied.
\\
The relatively thin glass used for the heliostat mirrors (4 mm thickness)
---leading to a low overall heat capacity---and the proximity of 
the ocean lead to frequent dew formation on the mirrors
in the winter.
To prevent micro drop formation, all mirrors were sprayed every second day
with a tensid solution in the evening using a specially constructed
spray cart. This procedure was found to work well after all mirrors
had been cleand with sulfonic acid from traces of silicon gel---a common
contaminant in glass production.

\subsection{Detector setup}

\subsubsection{Secondary optics}

\begin{figure}[ht]
\epsfig{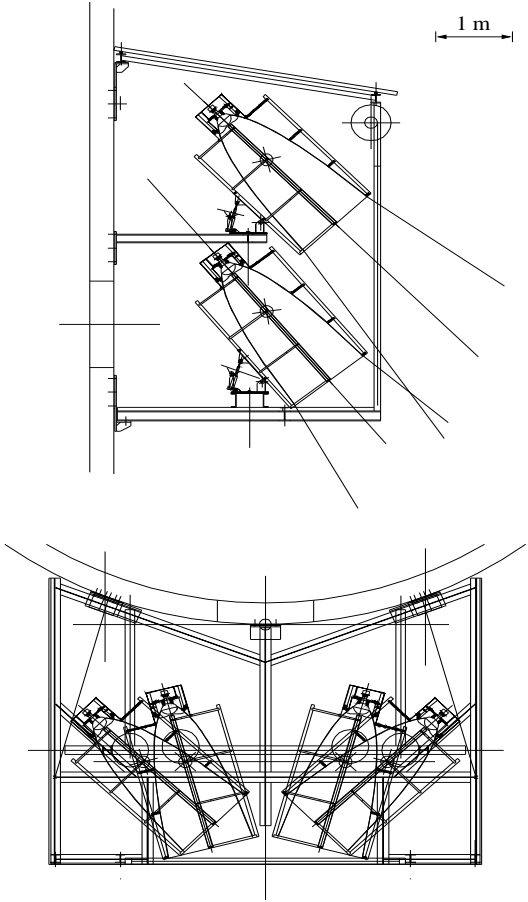}
\caption{\it Upper plot: Side view of the detector platform 
at the 70 m level of the central
tower. Two of the four Winston cones pointing towards their
respective heliostat subfields with the large-area
PMTs at the ends are sketched.
The wall of the central tower is at the left with
a manhole to enter the platform. Lower plot: View from above,
all four Cones are shown, the half circle is the wall of
the central tower.
}
\label{hut}
\end{figure}

\begin{figure}[ht]
\epsfig{
file=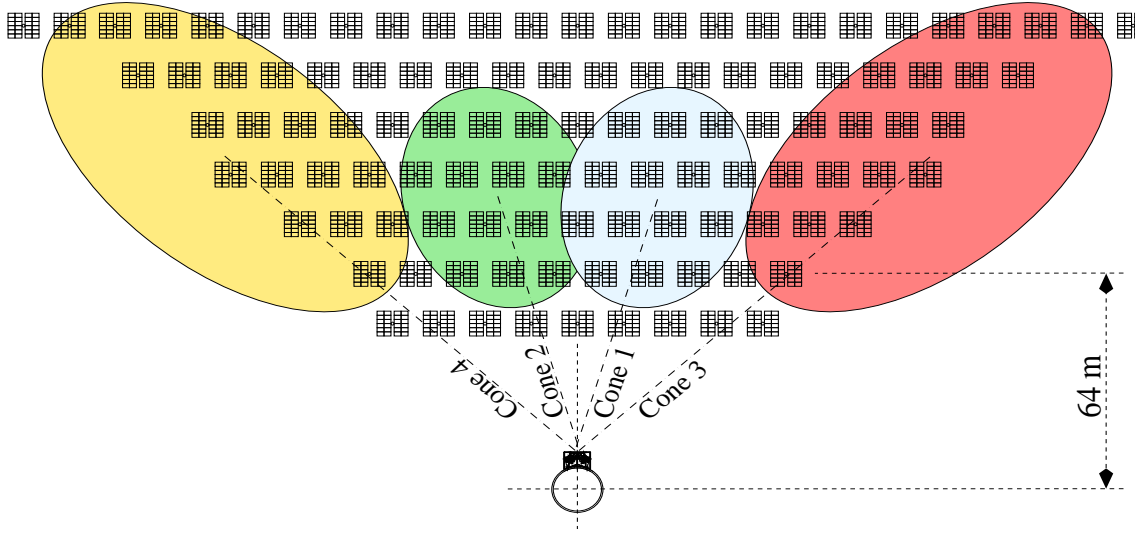,width=7cm,height=15cm,clip=,angle=90}
\caption{\it
Scheme of the detection geometry seen from above, north
is to the top of the page. The small circle
is the tower, the tiled double square symbolize the heliostats of CESA-1 
in the 2nd to 7th row from the tower.
The light from one of the four groups of heliostats used in GRAAL 
- indicated by the ellipses -
is concentrated into one of the four cones.
The cone numbering indicated is used throughout the text. 
}
\label{ground}
\end{figure}

\begin{figure}[ht]
\epsfig{
file=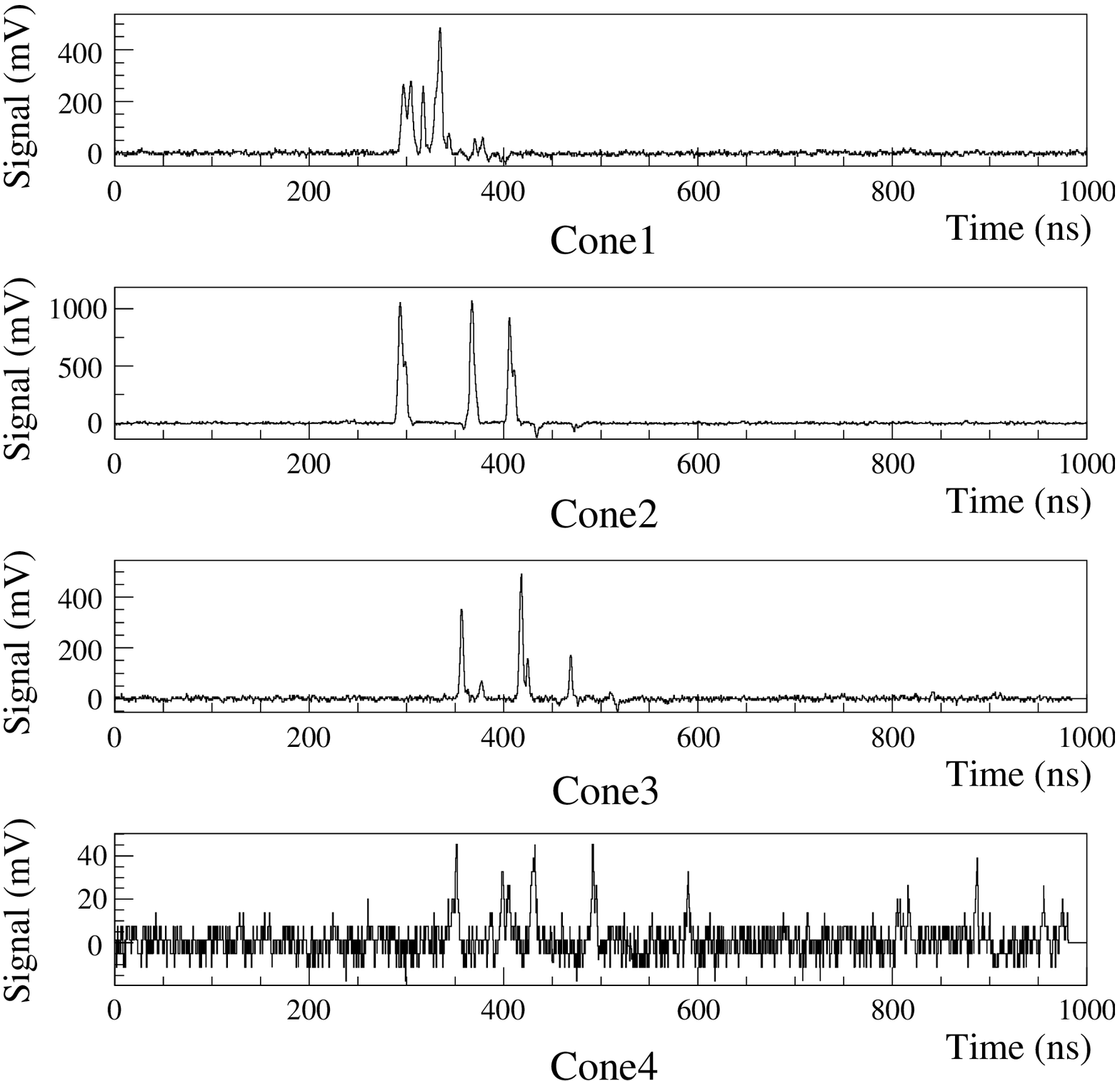,width=12cm,height=7cm,clip=,angle=0}
\caption{\it The signal height in mV after
amplification recorded in all four cones 
from one typical airshower is displayed as a function of time.
The trigger occurs at 500 ns. The y-gain depends on amplitude, at 100 mV
one mV corresponds to typically 0.25 photo electrons.   
Each peak corresponds
to the Cherenkov-light flash of the shower reflected by a different
heliostat. The distribution of light intensity on the ground within
the field of view of the cones is very uneven, note the different
y-scales. 
}
\label{exshower}
\end{figure}

Cherenkov light from four groups of heliostats  
(with 13,14,18,18 members, respectively)   
is directed onto four single non-imaging ``cone concentrators''  
each containing a single large-area photomultiplier tube (PMT).
The light collectors have the form of truncated Winston  
cones with an opening angle of 10$^{\circ}$. 
Each cone has a front diameter 1.08 m and a length of 2.0 m.
The cones are housed in a special enclosure that is fastened to the
outside of the central tower at the 70 m level (see fig.\ref{hut}).
Each cone is  
directed onto a point on the ground in the heliostat field  
and collects the light from all heliostats which are located  
within the ellipse projected by the cone opening angle  
on the ground (see fig. \ref{ground}).  
\\
At the end of each cone, a six-stage 8 inch hemispherical 
PMT, optimized for operation under high
background light levels (EMI 9352KB) is situated.
The tubes were typically operated with about 1300 - 1600 V
at a gain of about 8000. The signals were transmitted via
AC coupling to one fast amplifier directly adjacent to the PMT 
and a second one near the data acquisition electronics within the tower.
These amplifiers have a bandwidth of 
about 350 MHz and a gain of about 15 each. The final FWHM width
of Cherenkov pulses is about 3.6 ns, and mainly determined by
path length differences within the PMT.
\\
The incoming light from an air shower consists of a train of   
pulses from the different heliostats, usually fully separated  
by pathlength differences. The arrival time and amplitude
of each heliostat can thus be determined with a flash-ADC in
a sequential mode (fig.\ref{exshower}).

\subsubsection{Trigger logic}
\label{trigger}
Two completely independent triggers are used. For the ``sequence trigger''
after a discriminated signal above 30 mV a gate of 40 ns length
is opened after a delay of 20 ns. If a further signal is
detected during gate duration, another gate of 40 ns is opened with
a delay of 20 ns. If a third signal is detected in this second gate
an event-trigger gate of 200 ns is opened. If the first and second cone
have a coincident event-trigger the final event trigger is formed.
\\
For the ``charge(q) trigger'' a timing-amplifier integrates the
signal with an exponential time scale of 100 (200) ns for
Cone 1+2 (3+4). The integrated signal is fed into a discriminator in 
all four cones and opens a coincidence gate of 200 ns duration
if a preset threshold is surpassed (see also fig.\ref{ctrigger}, where
the Monte Carlo simulation of this trigger is discussed).
The singles rate of this integrated signal is the ``q-rate''
(table \ref{crabresult} - \ref{pseudoresult}).
A majority coincidence requirement of
``3 out of 4 cones'' is required for the final event trigger.
\\
Both triggers are always in a logical OR mode in data taking.
The event rate of the ``sequence trigger'' depends sensitively
on the incoming direction of the shower but is relatively insensitive
on the level of night-sky background induced background light
(NSB). The ``charge trigger'' is more strongly 
influenced by the NSB, but triggers on coincident signals independent
of detailed hypotheses on the arrival-time structure.

\subsubsection{Data readout} 

GRAAL achieves a good time resolution  
because there exist only four short cables   
that run exclusively within the  
platform enclosure from  
the photomultipliers to the data acquisition electronics. 
We register all four pulse trains  
in only one Digital Oscilloscope (Le Croy LC 564A)  
with a bandwidth of 1 GHz and a time bin of 500 ps.  
This ensures that the FWHM of individual pulses of about 3.6 ns 
is negligibly increased by electronics effects. The digital scope is  
read out in sequence mode over a GPIB interface into a PC, reaching  
a speed of about 260 ``waveforms''/sec  
(i.e. 1000 time bins of 0.5 ns width with 1 byte each),  
which is sufficient for a dead time below 10 $\%$ for  
our master trigger rate which always remains below 5 Hz
and is typically about 2-3 Hz (each trigger containing  
four waveforms). 

\subsubsection{Calibration}
\label{cal}
The time and amplitude calibration of our setup  
is performed using blue LEDs (Nichia NSPB 500, maximal  
output at 470 nm)   
with a calibrator  
module that is fastened at the window of the Winston cones.  
The amount of light emitted   
by the individual LEDs is determined with a Quantacon RCA C31000
(a photomuliplier yielding a well separated single photoelectron
(p.e.) peak)  
that was previously calibrated by determining   
its single p.e. peak and fluctuation behaviour.  
The LED operating voltage is adjusted so that  
one LED pulse corresponds to about 100 p.e.  
These LED pulses are regularly used in each run  
to verify time and amplitude calibration.   
In addition a LED module with higher
total light output shines onto the heliostat   
field. When the heliostats are brought into  
a ``back reflection'' position, the reflected  
LED pulses are used to verify the geometry and  
check the mirror quality.
\\
The timing and gain properties of the electronics chain were
calibrated on-line with a PC-controlled Phillips Scientific PS7120 
charge injection module. Charge pulses with properties similar
to PMT pulses and different amplitudes 
were injected directly after the PMT in each run.

\subsubsection{Remote operation}
All operations (like opening of the door, high-voltage control etc.)
at the central receiver and the tracking of the heliostat field
are under remote control 
via the internet. Various
environmental parameters like humidity, ambient light, wind speed, rates etc. 
are checked by the data-acquisition computer. Under conditions
that indicate some malfunction, a physicist on shift is phoned by the PC
and can check all parameters and images of web cameras, remotely.
For the operation of the heliostat field and emergencies only the regular
night-operator of the PSA is on-site in all observation nights.

\subsection{Differences of the 
basic ``CELESTE'' versus ``GRAAL'' central-receiver approach}
\label{diff_cel}
After early tests for the use of heliostat fields
for $\gamma$-ray astronomy \cite{ori} the basic idea of
T\"umer\cite{tuemer1,tuemer2} to image the Cherenkov light of one heliostat
to a single photomultiplier has been worked out in technical
detail in the proposal for the CELESTE experiment\cite{celesteprop}.
It was then proven technically at the Themis heliostat array
\cite{celestetech}. 
Two other heliostat-field experiments 
``STACEE''\cite{staceeprop,staceetech} and
``Solar 2''\cite{solar2} follow this basic design.
Recently CELESTE\cite{celestecrab} and STACEE\cite{staceecrab} 
reported the detection of
VHE $\gamma$ rays.
The major differences between this well documented method
to detect air showers with heliostat arrays
and the ``non-imaging'' principle of GRAAL---which
collects the light from 13-18 heliostats in a Winston cone
onto a {\it single} large-area photomultiplier---are described in the following.
\\
$\bullet$
The most important drawback of the non-imaging approach of GRAAL
is that the night-sky background is higher roughly by the number
of heliostats viewed by one cone.  This results in a typical expected
background of 8-10 p.e./ns in GRAAL, compared to
0.7 p.e./ns in CELESTE. The hardware
energy threshold for the detection of $\gamma$-rays in principle
achievable with the same mirror area used is about 4 times higher
in GRAAL. For pulses far above threshold
the performance of the two approaches is not expected to be very
different because a similar amount of Cherenkov
light is gathered by GRAAL and CELESTE.
\\
$\bullet$
The advantage of the non-imaging approach is its greater simplicity
leading to savings by about a factor 5-10 in hardware costs.
The presence of only  four data-acquisition 
channels makes automatization and
remote control more  feasible, leading to comparable
savings in operation costs.
In its present configuration GRAAL normally runs under remote
control with only a PSA operator (who is present for maintenance of the
facilities independently of GRAAL) on-site. 
The small number of channels allows to use flash-ADCs with a
time resolution of 0.5 ns/bin, higher than any other 
Cherenkov experiment.
\\
$\bullet$ 
In CELESTE
the angular field-of-view in the sky of each PMT
is designed to be
constant at 10 mrad (full angle). In GRAAL this is impossible
because the contributing heliostats' distance from the collecting cone
varies.
This field-of-view therefore varies between 6.5 and 12.1 mrad.
It is not easy to determine
the ``optimum'' value for the field of view since it depends on several diverse
factors. The total acceptance has to
be derived from detailed Monte Carlo simulations even
in case of a fixed acceptance. Therefore this difference
seems of little importance.
\\
$\bullet$ 
Because the non-imaging approach of GRAAL requires
that groups of directly adjacent heliostats in the fields
are chosen, its configuration is more compact. In GRAAL 63
heliostats that cover an area of about 160 $\times$ 80 m$^2$ are used, 
whereas CELESTE presently uses 40 heliostats that cover an area 
of 240 $\times$ 200 m$^2$, i.e. the sampling density is about a factor 5 lower.
From the Monte-Carlo simulations it seems that
with a restricted field of view 
the irregular structure of the light pool in hadronic showers
tends to be more pronounced 
at large distance scales, so a more extended array tends to be 
advantageous for a possible $\gamma$-hadron separation.
\\
$\bullet$
In the non-imaging approach it is impossible to avoid a temporal
overlap of the signal from certain heliostats depending on the
pointing direction. This reduces the number of times/amplitudes
usable in the reconstruction by about 20$\%$. When the incident
direction lies northward (this is the case for the source 3EG 1835+35
at the location of GRAAL),
the overlap becomes stronger leading
to a substantial decrease in the quality of reconstruction.
On the positive side, calibration is easier when signals from several
heliostats are measured in the same PMT.
\begin{figure}[ht]
\epsfig{
file=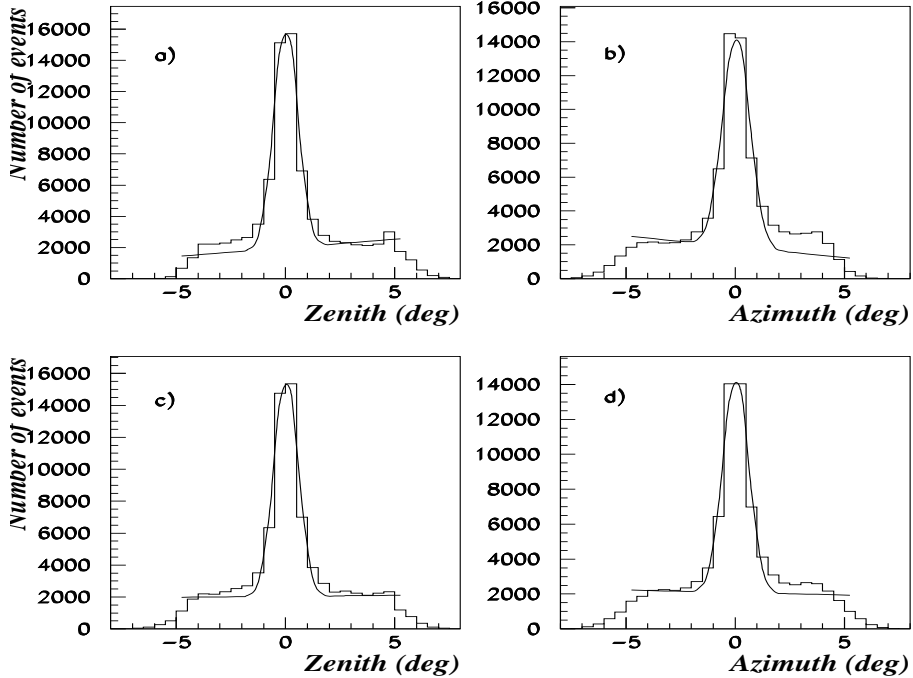,width=12cm,height=9cm,clip=,angle=0}
\caption{\it Projections of the number of showers as a function
of shower directions as reconstructed from the timing data. 
Shown is deviation of the reconstructed direction 
from the pointing direction 
on the elevation-axis (left two panels a. and c.) and azimuth-axis 
(right two panels b. and d.). 
The origin then corresponds to the pointing
direction as determined by the orientation of the heliostats.
Two components are apparent: a peak near the origin, 
and a ``flat background'' corresponding
to events misreconstructed in direction (see text).
The data sample comprises
of 32 hours of ON-source time on the Crab pulsar (upper panels a. and b.) 
and an equal amount of
OFF-source time (lower panels c. and d.) taken under variable 
weather conditions in the season
1999/2000. The ``Gaussian plus linear function'' fit 
is performed to each subsample.
It is seen that the Gaussian - correspondig to successfully 
direction reconstructed events -
is always centred within $<$ 0.05$^{\circ}$.} 
\label{angproj}
\end{figure}
\section{Event reconstruction}
\label{evreco}
\subsection{Software-trigger threshold}
\label{evreco_tri}
The night-sky background (NSB) RMS fluctuation was estimated from
a portion of the flash-ADC recorded traces that do not 
contain Cherenkov signals,
for each event and cone individually.
The arrival time of all detected signals with an amplitude exceeding
n$_t$ $\times$ $\sigma_{NSB}$ was determined from the 
recorded full pulse shape 
in the related flash-ADC. 
The parameter n$_t$ was chosen to be typically between 5 and 7.
These arrival times were corrected for path length differences in
cables and within PMTs with the online-calibration (section \ref{cal})
and were then used to reconstruct the timing shower front of
the individual events. Arrival times closer to each other than 6 ns
were excluded to avoid any bias from overlapping pulses. Signals
that saturated any channel were also excluded from further analysis.
Only the NREMAIN remaining signals
were used in the further analysis.
\\
Before further analysis a software-trigger threshold was applied.
In order to allow a meaningful reconstruction of shower parameters
NREMAIN $\ge$ 5 was required.
n$_t$ was chosen at a value as low as possible, before a large
number of NSB induced ``fake'' signals were found to enter the sample.
The lowest possible value of n$_t$ was found to depend on the source position
somewhat, due to the varying temporal overlap of signals in the trace.
The final choice was: n$_t$=5(7,9,7)
for the Crab (3C454, 3EG+1835, pseudo source) sample.
This software threshold also equalized the effect of the NSB on the
reconstruction. A higher level of NSB $\sigma_{NSB}$
leads to a correspondingly higher software threshold. This is
expected to correct for the effect of a lower hardware trigger threshold
and decreased reconstruction efficiency
with higher NSB to first order.
 The choice of n$_t$ in the analysis of MC data was different and is
 explained in section \ref{effdet}.

\subsection{Reconstruction of incoming shower direction}
\label{evreco_dir}
\begin{figure}[ht]
\epsfig{
file=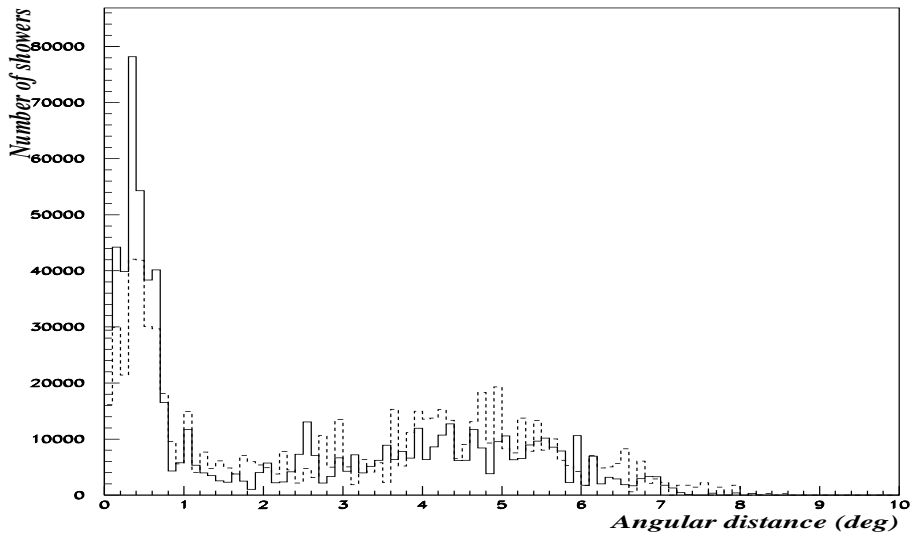,width=12cm,height=7cm,clip=,angle=0}
\caption{\it 
Monte Carlo simulation of angular reconstruction of events
from a $\gamma$-ray point source (full line, zenith angle 
10$^{\circ}$, azimuth angle
45$^{\circ}$) and diffuse source of protons (dashed line). 
Shown is the number of showers
as function of angular distance from the pointing direction
in degrees. It is seen that the relative fraction of showers
with ``misreconstructed directions'' of the total data sample
(``flat background'' in fig.\ref{angproj}) is 
much larger for protons (see text).
The ratio r$_{io}$ = (``events with angular 
deviation $<$ 0.7 degree''/ ``all events'') 
is 0.45 and 0.29 for gammas and protons, respectively.} 
\label{angdist_mc}
\end{figure}

The expected arrival times for all heliostats in each of the
four cones were calculated and stored in a ``library''
for a 5 $\times$ 5 degree grid
centred to a direction about 1 degree offset from the current
pointing direction of the heliostats. The offset was chosen to
avoid a bias towards ``correct pointing''.
This calculation was performed assuming a point-like
shower-maximum at a penetration depth of 230 g/cm$^2$
(the mean penetration of showers induced by a photon of 100 GeV)
in the pointing direction. A spherical timing-front was assumed
to be emitted by this maximum. Tests with plane and parabolical
timing fronts showed, that while the former leads to 
worse fits to the timing data, the latter does not improve the quality
of the fit significantly.
The shower core was fixed at the
geometrical centre of the field as defined by the heliostats used.
\\
We attempted to reconstruct the position of the 
shower-cores of individual showers on the ground 
using the recorded amplitude information.
Different light-gathering efficiencies of heliostats
due to different distances to the tower, mirror quality etc.
were corrected via normalizing the amplitudes over many showers
and then the centre-of-gravity of the light distribution was
determined.
It was verified that the mean of the centre-of-gravity over all detected
showers lies at the geometrical centre of the field used within 1 m
so that the assumption of a ``fixed core'' 
at this position introduces no bias. 
From the Monte-Carlo data it was found that---due to
the rather compact size of our field---a shower core reconstructed
for each individual shower from the amplitude information 
has a larger mean deviation from the true
core location than the ``fixed core''.
Therefore we assumed that all shower cores
lie at the "fixed core" in the
reconstruction algorithm. 
\\ 
The measured arrival
times were compared to this ``library''. We define the time difference
TIMEDIFF
\begin{equation}
\rm{TIMEDIFF=(measured\: arrival\: time)-(nearest\: expected\:
time\: from\: the\: library)}
\label{timediff}
\end{equation}
The direction yielding the smallest ``${\rm lsq_t^2}$''
defined as the least squares sum:
\begin{equation}
{\rm lsq_t^2} =\sum_i ({\rm TIMEDIFF}_i)^2 
\label{chi2}
\end{equation}
was chosen as the final 
reconstructed direction of the shower. 
\\
There is a possibility that spurious pulses induced 
by the night-sky background,
after pulsing in the PMTs or due to cross talk between the subfields occur.
These pulses do not fit into the correct timing pattern and bias the fit.
Up to n TIMEDIFFs above 5 ns
were therefore allowed not to be taken into account in the calculation
of the ${\rm lsq_t^2}$ (n=5 was chosen for all analyses discussed
in this paper).
\\
This procedure was performed on a grid of 0.5 degree step width, the final
direction was improved via a quadratic fit to ${\rm lsq_t^2}$ values of
the four grid points adjacent to the one with smallest ${\rm lsq_t^2}$.
Fig.\ref{angproj} shows projections of reconstructed directions
in zenith and azimuth angle both for ON and OFF source directions
for a large data sample. The origin corresponds to the
pointing direction determined by the heliostat tracking.
A combined fit is performed with a Gaussian for the events reconstructed
near the centre and a linear function for the 
``smooth background'' extending to
large off-axis angles.
\\
The directions of events in this ``smooth
background'' were found to be systematically
misreconstructed. The ${\rm lsq_t^2}$ of the timing
fit of these events is found to be systematically {\it lower} than for
the ``central'' events because the incorrect reconstructed direction
allowed incorrect ``heliostat - measured signal'' assignments.
Fig.\ref{angdist_mc} compares the reconstructions for gammas and
hadrons. It is seen that proton induced showers - presumably because
of a systematically higher fluctuation in arrival times - are more
prone to this misreconstruction effect and therefore populate the
``smooth background'' preferentially. This 
effect is used in the later analysis (section \ref{datared})
to normalize the ON and OFF rates.
\\
If the ``misreconstructed'' directions are excluded, the
angular resolution $\sigma_{63}$ (the opening angle within which
63$\%$ of all events are contained) is 0.7$^{\circ}$.

\section{Monte Carlo simulation of experiment}
\label{moncar}
Proton and $\gamma$-induced showers were simulated with the
Monte-Carlo package Corsika 5.20\cite{corsika}. A detector Monte Carlo
simulated the reflection of the Cherenkov photons by the
heliostats mirrors into the cones and the further
processing of the signals in the PMTs and electronics.
The heliostats were approximated as spherical
mirrors in two rectangular sections of 6.75 $\times$ 3 m$^2$.
Imperfections of the surface were simulated to give results in accordance with
results on imaging of the sun onto a screen below the GRAAL receivers.
PMT properties were simulated according
to manufacturer specification.
The properties of electronic components (amplifiers, active splitters etc.)
were deduced from the charge injection measurements (section \ref{cal}).
A careful modelling was necessary here, because for example our 
fast preamplifiers have a strong nonlinearity, in particular their
gain rises with amplitude and frequency (up to
about 300 MHz).

\subsection{Monte-Carlo data generation}

All Monte-Carlo results reported in the present paper were obtained
from a library of simulated showers initiated by both gamma-rays 
and protons. Primary energy ranged from 0.05 to 1 TeV for gamma-rays 
and from 0.2 to 4 TeV for protons. For both primaries the events were 
generated with a differential energy spectrum following a power law 
with index -1 instead of the real one. This procedure allows 
a reasonable statistics to be attained at high energy without having to
produce a non-affordable number of events at low energy. For the
calculation of all the detector parameters each shower was assigned a 
weighting factor in such a way that the corrected spectral index for 
protons was -2.7 and that of gamma-rays was -2.4. 
While gamma-rays were generated as
incident from a point-like source in the observed direction, the
incoming directions of protons were randomly generated around the
observed direction with a maximum angular deviation of 4 degrees.
The core position of the showers was randomly generated up to a
maximum distance from the centre of the array of 300 m. For
computing time reasons r has followed a probability law P(r)dr = Cdr 
and consequently the events have been assigned a weight proportional to r.
As a further procedure which has been employed 
to maximize the usefulness of the CPU time, for every simulated shower 
the GRAAL response has been calculated for 5 different core positions
following the above law.
8000 independent showers for each species were simulated for each of 6
incident directions.

\subsection{Simulation of the night-sky background (NSB)}

The signal generated by each shower in the four PMTs has been stored in
histograms with a bin size of 0.5 ns. 
The NSB contribution has been added by injecting for each bin a signal
waveform whose amplitude corresponds to a Poisson distributed number of 
photo-electrons.
The mean value of this distribution (about 6 p.e.) was calculated 
for each PMT, adding up the NSB contribution of 
every heliostat with the detailed ray-tracing algorithm of the detector 
simulation using the known brightness on small angular 
scale of the night sky at the location
of Almeria (1.8 $\times$ 10$^{12}$ photons m$^{-2}$ sec$^{-1}$ sr$^{-1}$
inferred from the wide-angle value measured by Plaga et al.\cite{plaga95}).
For every simulated shower (and core position) 
a new (statistically independent) NSB has been generated.

\subsection{Effective detection area}
\label{effdet}
\begin{figure}[ht]
\epsfig{
file=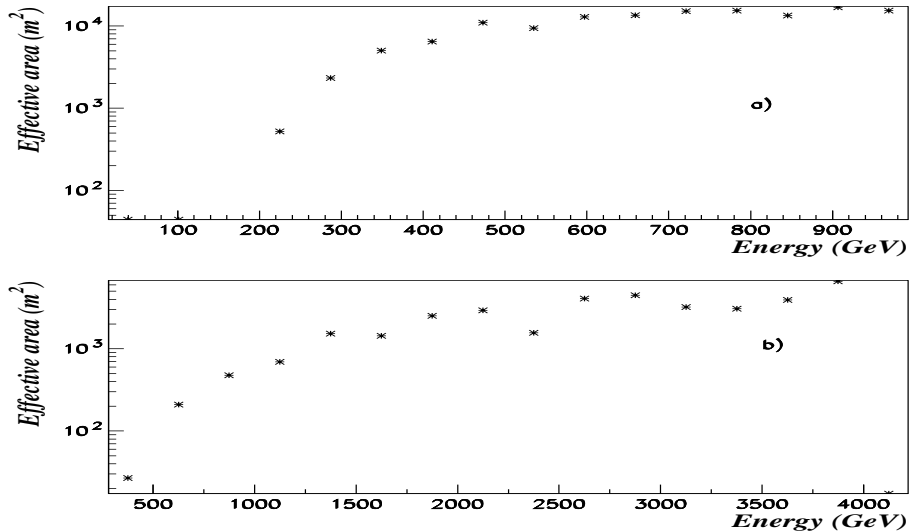,width=12cm,height=7cm,clip=,angle=0}
\caption{\it Effective detection area for gammas (upper panel a.) and
protons (lower panel b.) incoming from a zenith angle of 30 degrees and
azimuth angle of 0 degrees.}
\label{detarea}
\end{figure}

The weighted Monte-Carlo data sample was used to estimate the
effective detection area for protons and gamma-rays at a zenith angle of
30$^{\circ}$ and azimuth angle of 0$^{\circ}$ 
as a function of primary energy
(fig. \ref{detarea}). (The azimuth is counted from south towards the east
in this paper). 
All showers passing the ``software-trigger 
threshold'' in the real data
as defined in section \ref{evreco_tri} were counted as detected in 
this simulation.
\\
We chose a parameter n$_t$=9 
of section \ref{evreco_tri} 
(software-trigger threshold for single signal in units
of night-sky background RMS),
to obtain a proton induced
rate of 4 Hz, in agreement with the typical 
experimentally observed value.
Fig.\ref{detarea} was obtained with this value for n$_t$.
The chosen value of n$_t$ is somewhat higher than the
one used for the experimental data (n$_t$ = 5-9
(section \ref{evreco_tri})). This would imply that the
experimental signals are somewhat smaller
than the ones predicted by the MC simulation, relative
to the level of the NSB.
This is the reason why the threshold
of GRAAL is higher than originally expected\cite{kruger}.
The disagreement between data and MC is most likely 
due to a combination of factors
not taken into account in the Monte-Carlo simulation, like after-pulsing
in the PMTs, a loss in light-collection efficiency due to
in-operational heliostats, dust on the heliostat surfaces and in the air
and imperfect canting conditions. 
A more precise procedure where the hardware trigger is simulated in
detail gave similar results\cite{davidthes}.
\\ 
Because gamma-ray and proton induced showers
were found to be very similar (see e.g. fig.\ref{timed}) 
$\gamma$-ray fluxes determined 
relative to proton fluxes (section \ref{datared}) are 
expected to be biased negligibly
by this somewhat unrealistic trigger simulation. 
Results obtained for a zenith
angle of 10 degrees and azimuth angle of 45 degrees were very similar.
The effective energy threshold for gamma-rays, defined as the 
maximum in a plot of differential
flux versus primary energy, derived from the 
panel a. of fig.\ref{detarea} is
250 $\pm$ 110 GeV at 10$^{\circ}$ zenith angle and 300 $\pm$ 130 GeV
at 30$^{\circ}$ zenith angle.

The areas in these diagrams were used to caluculate the expected total
event-trigger rate.   As stated above, the parameter $n_t$ was adjusted such
that the resulting effective area for protons, combined with the known
absolute differential flux of cosmic-ray protons $\phi_{\mathrm ref}$\cite{wiebel},
yielded a proton-induced  shower rate of $r_p = 4.0\:{\mathrm Hz}$,
corresponding to the measured value. The differential $\gamma$-ray flux from the
Crab nebula above 500~GeV as determined by the Whipple collaboration\cite{whipple} was
used to estimate the gamma-induced shower rate of 
$r_\gamma = 0.011\:{\mathrm Hz}$.
\subsection{Comparison of Monte-Carlo generated with experimental data}
\label{compar}
We compare the distribution of some basic measured parameters for 
protons and gammas simulated for
an incident zenith angle of 30 degrees and azimuth
angle of 0 degrees with data taken in the zenith angle range
25 - 35 degrees and an azimuth angle between 310 and 322 degrees.
The threshold parameter n$_t$ (section \ref{evreco_tri}) was set to
6 for the Monte-Carlo data used for the construction of figures
\ref{numberpeak} - \ref{totchar}.
This is slightly smaller
than the value of 7 which was chosen for the comparison experimental data 
(taken on the source 3C 454.3 in all
figures). The motivation for this decrease is that
- as explained in section \ref{effdet} - the experimental
signals seem to be somewhat smaller than expected from the
MC simulation.
Some parameters of the reconstruction procedure
were found to depend quite sensitively on the ratio signal/NSB
in the Monte Carlo simualtions.
We chose n$_t$=6 for the MC simulations in order to reproduce correctly the
experimentally observed ratio
r$_{io}$ as defined in figure 6.

\subsubsection{Number of heliostats with detected signal}
\begin{figure}[ht]
\epsfig{
file=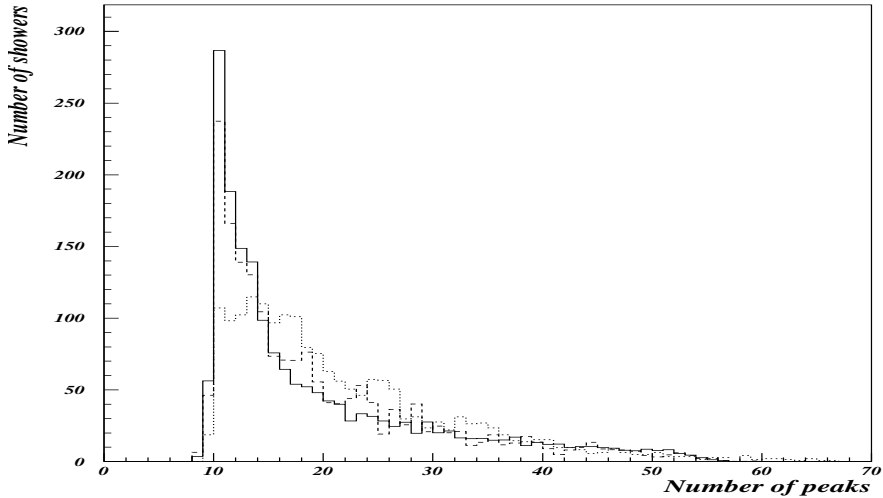,width=12cm,height=7cm,clip=,angle=0}
\caption{\it
Number of showers with a given number of peaks identified in all
four recorded timing traces. 
The full (dashed) line is for MC $\gamma$s (protons), and the dotted line for
experimental data taken under similar incident angles.
The total number of MC showers was normalized to the experimental data for
comparison.
}
\label{numberpeak}
\end{figure}

\begin{figure}[ht]
\epsfig{
file=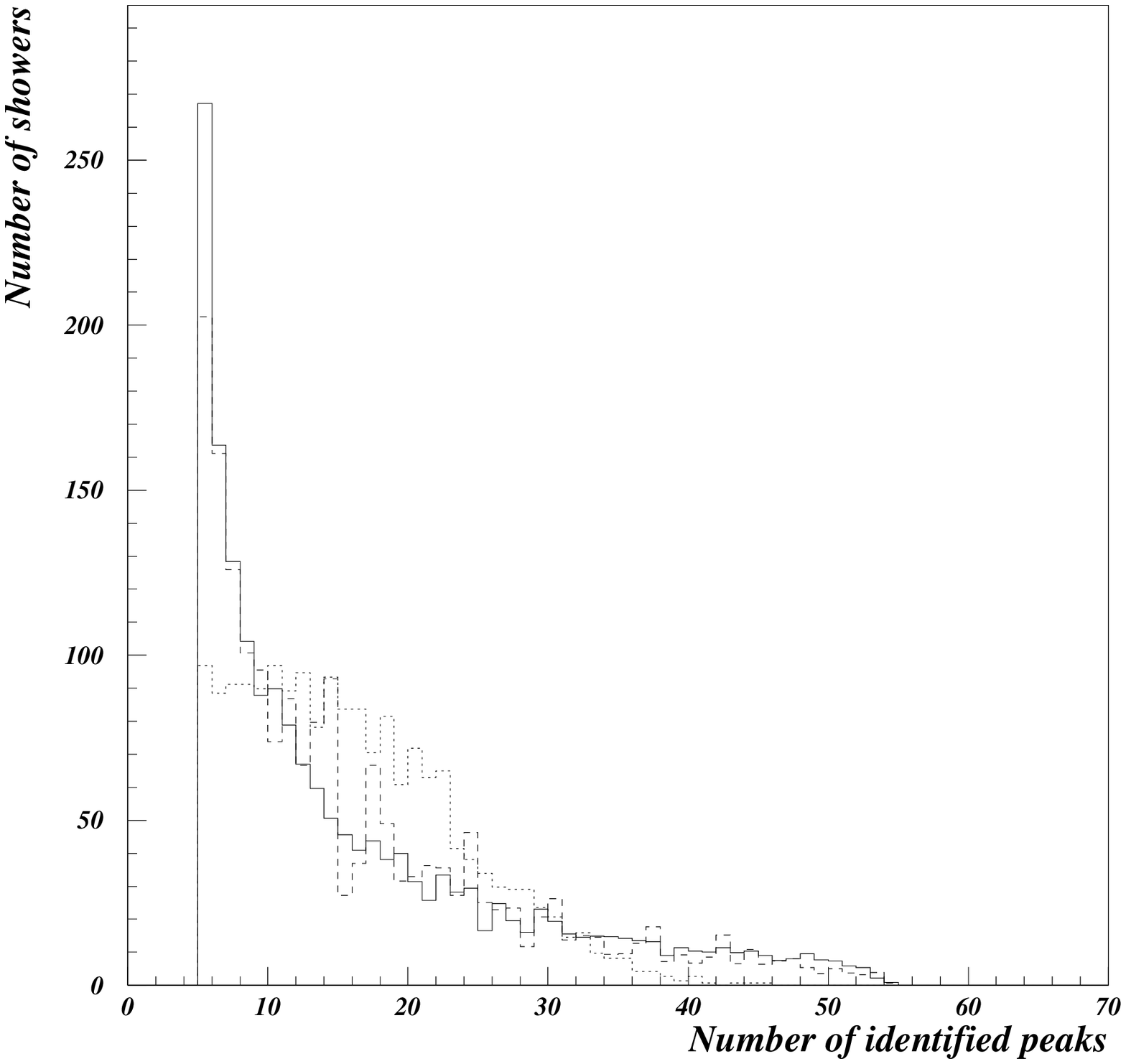,width=12cm,height=7cm,clip=,angle=0}
\caption{\it
Number of showers with a given number of peaks that were attached
to individual heliostats and were used in the final determination
of the shower direction.
The full (dashed) line is for MC $\gamma$s (protons), and the dotted line for
experimental data taken under similar incident angles.
The total number of MC showers was normalized to the experimental data for
comparison.
}
\label{inumberpeak}
\end{figure}

A basic parameter is the number of Cherenkov flashes from individual
heliostats that have been identified
and are the input values for the
reconstruction of the shower timing-front (called
NREMAIN section \ref{evreco}).
Fig.\ref{numberpeak} shows the distributions of NREMAIN.
Some peaks cannot be identified as being due to a reflection
from a given heliostat and are not used for the reconstruction of
the timing front.
The mean (RMS) of the distribution for proton MC is 19.6 (10.0) 
and for the experimental data 21.7 (10.3).
Fig.\ref{inumberpeak} shows the distribution of the ``remaining''
identified peaks that could be attached to individual heliostats
and are actually used in the timing fit.
The mean (RMS) of the distribution for proton MC is 16.3 (10.9) 
and for the experimental data 16.0 (7.4).
From this, the fraction of identified peaks is 83$\%$ for protons
in the Monte Carlo and 73$\%$ in the experimental data.
The results of a $\chi^2$ test of the compatability of simulated
$\gamma$-showers and experimental showers to
simulated proton-induced showers
are reported in table \ref{chi2t}.
\begin{table}[ht]
\vspace{-10pt}
\caption{\it Results of a comparison of the distributions
of MC $\gamma$ versus MC proton-induced showers
in fig.\ref{numberpeak} - \ref{chic}.  $\chi^2_{\rm red}$($\gamma$/p) lists the
values from a comparison of $\gamma$ versus proton induced showers,
and  $\chi^2_{\rm red}$(data/p) a comparison of proton induced showers
and data.
$\chi^2_{\rm red}$ values that are acceptable
on the 90$\%$ level for the given number of degrees of freedom
n$_{\rm dof}$ are bold faced.}
\label{chi2t}
\begin{center}
\begin{tabular}{lccc}
 &   $\chi^2_{\rm red}$($\gamma$/p) &  $\chi^2_{\rm red}$(data/p)  &  n$_{\rm dof}$   \\
\hline
\hline
Total number of peaks (fig.\ref{numberpeak})   & {\bf 1.05}  & 3.09  & 70  \\ 
Selected number of peaks (fig.\ref{inumberpeak}) & {\bf 1.2}   & 4.67   & 70  \\ 
Squared time deviation (fig.\ref{chic})  & 1.54  & 1.72  & 200  \\
 
\hline
\end{tabular}
\end{center}
\end{table}

For the comparisons related to the 
number of peaks (figs.\ref{numberpeak},\ref{inumberpeak})
the $\chi^2$ values are compatible with identity of the 
distributions of MC protons and MC gammas
for the number of degrees of freedom, but the proton- and data distribution
differ significantly. This difference is due in both cases to
a disagreement near threshold and for very large showers,
whereas for the majority of intermediate showers
- with a number of peaks between about 15 and 40 - 
the agreement is satisfactory. 
The reason for the discrepancy for very small showers is 
probably that the discrepancy between data and MC
in the ratio if shower sizes and size of the NSB discussed
in section \ref{effdet} is not completely resolved by the choice
of slightly higher n$_t$ discussed in section \ref{compar}.
The discrepancy for very large showers is very likely due
to the fact that at each time typically about 10 heliostats were
inoperational. 
\\
In the {\rm lsq$^2_t$}
distribution, all $\chi^2$ values are incompatible with
identical parent distributions. While a difference between gamma-rays
and protons is expected (the former have slightly smaller deviations
from a spherical front), the large deviation of experimental showers
from a spherical front is very likely due to the effects of after pulsing
(introducing additional peaks with large time deviations).

\subsubsection{Timing properties}
\begin{figure}[ht]
\epsfig{
file=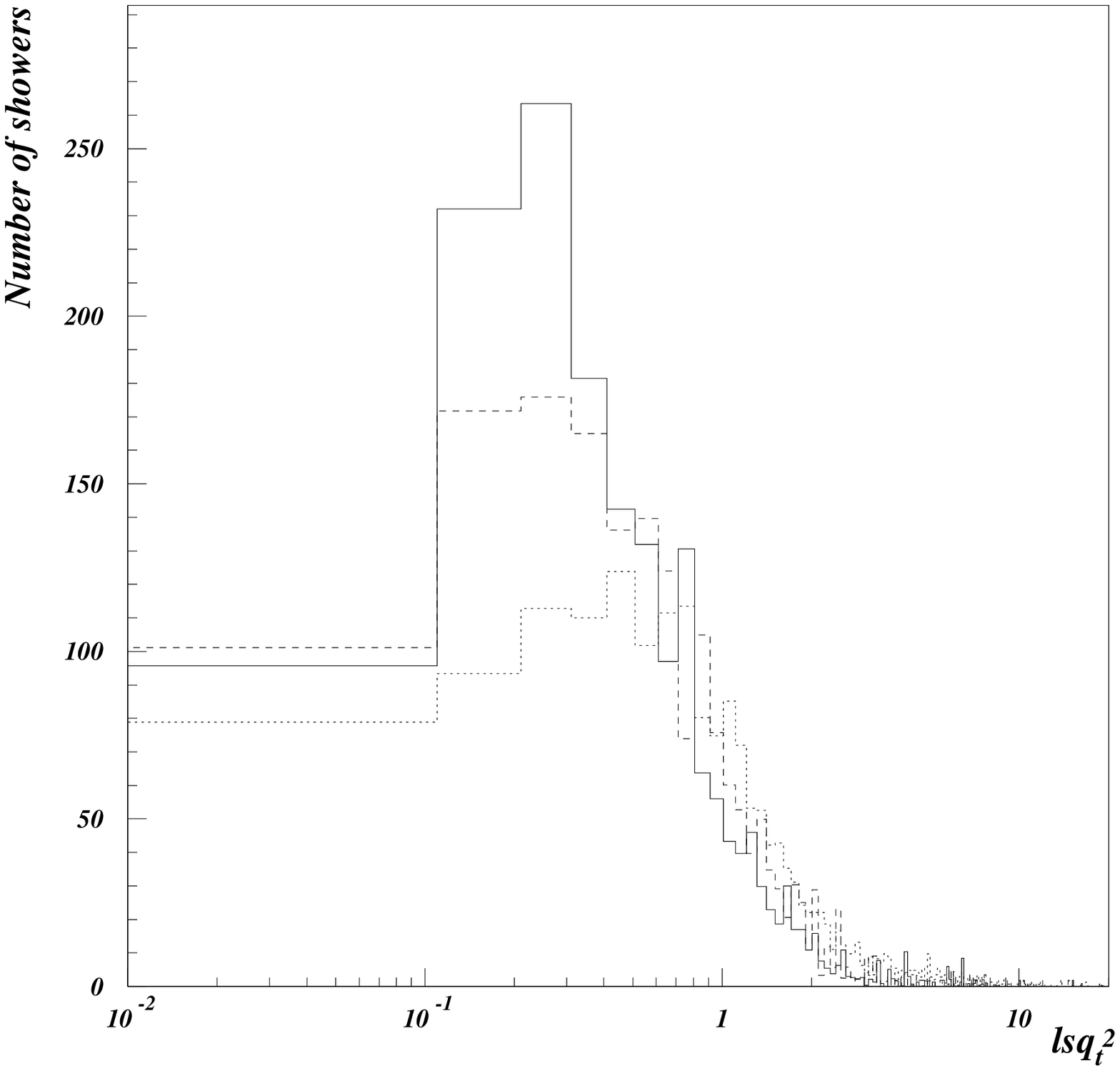,width=12cm,height=7cm,clip=,angle=0} 
\caption{\it 
The distribution of ${\rm lsq_t^2}$.
The full (dashed) line is for MC $\gamma$s (protons), and the dotted line for
experimental data taken under similar incident angles.
The total number of MC showers was normalized to the experimental data for
comparison.
}
\label{chic}
\end{figure}

\begin{figure}[ht]
\epsfig{
file=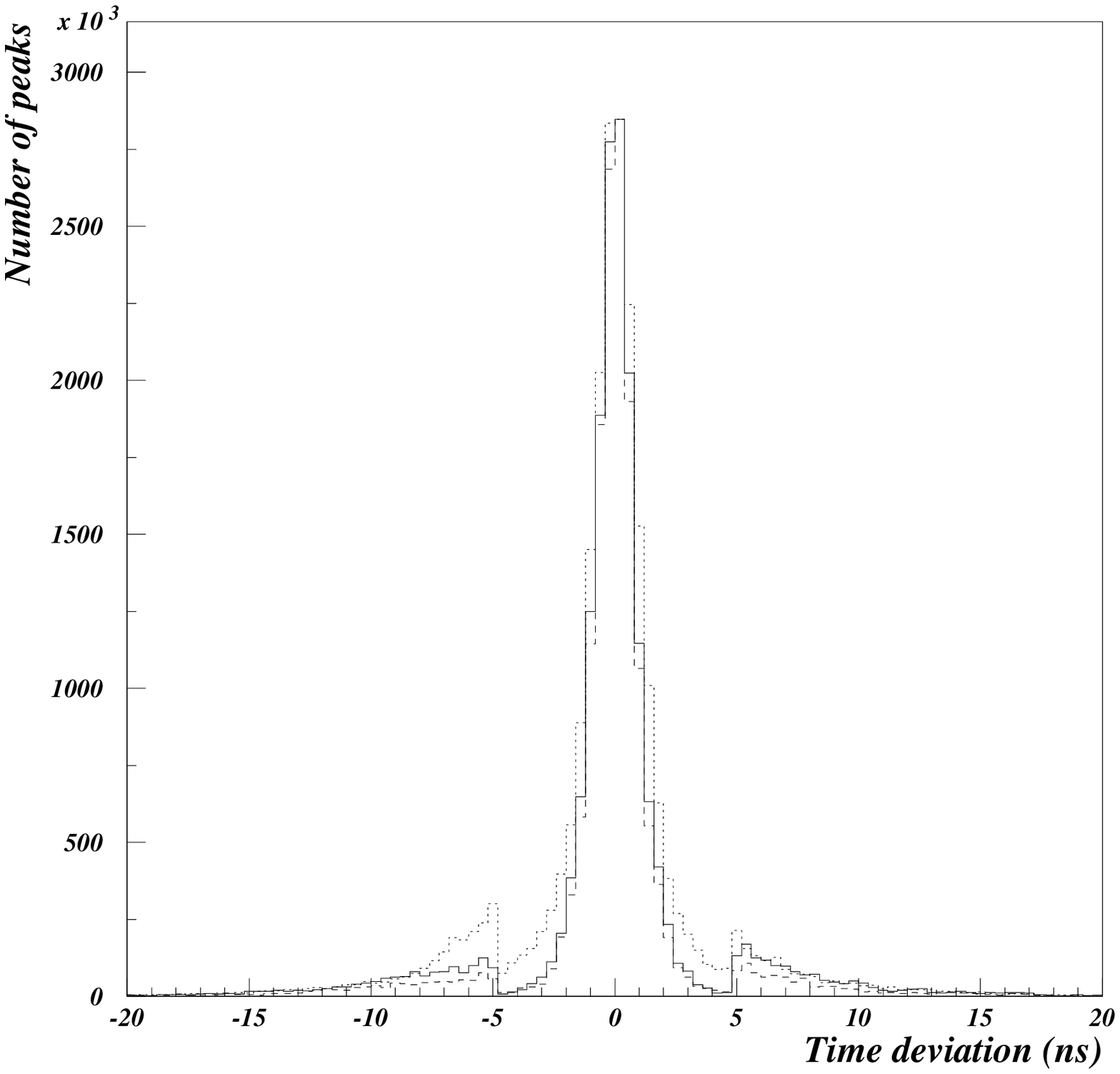,width=12cm,height=7cm,clip=,angle=0}
\caption{\it The deviation of measured arrival times from the final fitted
spherical shower front for MC $\gamma$´s (full), protons (dashed) and
experimental data (dotted). The visible sharp reduction of events with
a time deviation somewhat smaller than 
5 ns is due to the fact that the reconstruction
program allows the exclusion of 3-5 peaks with a deviation from the shower
front larger than 5 ns (see text) from the final fit.
}
\label{timed}
\end{figure}

The distribution of the ${\rm lsq_t^2}$ (eq. \ref{chi2}) of the timing fit
for MC simulated and experimental showers is shown in fig.\ref{chic}.
The distributions for MC simulated gamma´s and protons are remarkably
similar. The experimental data show a distribution with a mean which
is a factor 2.2 larger and a RMS which is a factor 1.4 larger than the
MC sample with proton-induced showers.
To understand the similarity of $\gamma$ and proton showers better,
fig.\ref{timed} shows the deviation of pulse arrival time from the
final spherical shower front for the optimal fitted direction.
One notices that the central peak has practically the same width
in Monte Carlo ($\approx$ 0.8 ns in Monte Carlo and 
1.0 ns in the experimental data). The times are very
close to the theoretical sphere, this is 
also true for simulated proton and experimental
showers. At large time deviations experimental data have many more
hits than MC data, presumably mainly due to after pulsing in the PMTs which
was not simulated in the MC. It is mainly these tails that increase the
mean of the experimental ${\rm lsq_t^2}$ distribution.

\subsubsection{Total-charge spectrum}
\begin{figure}[ht]
\epsfig{
file=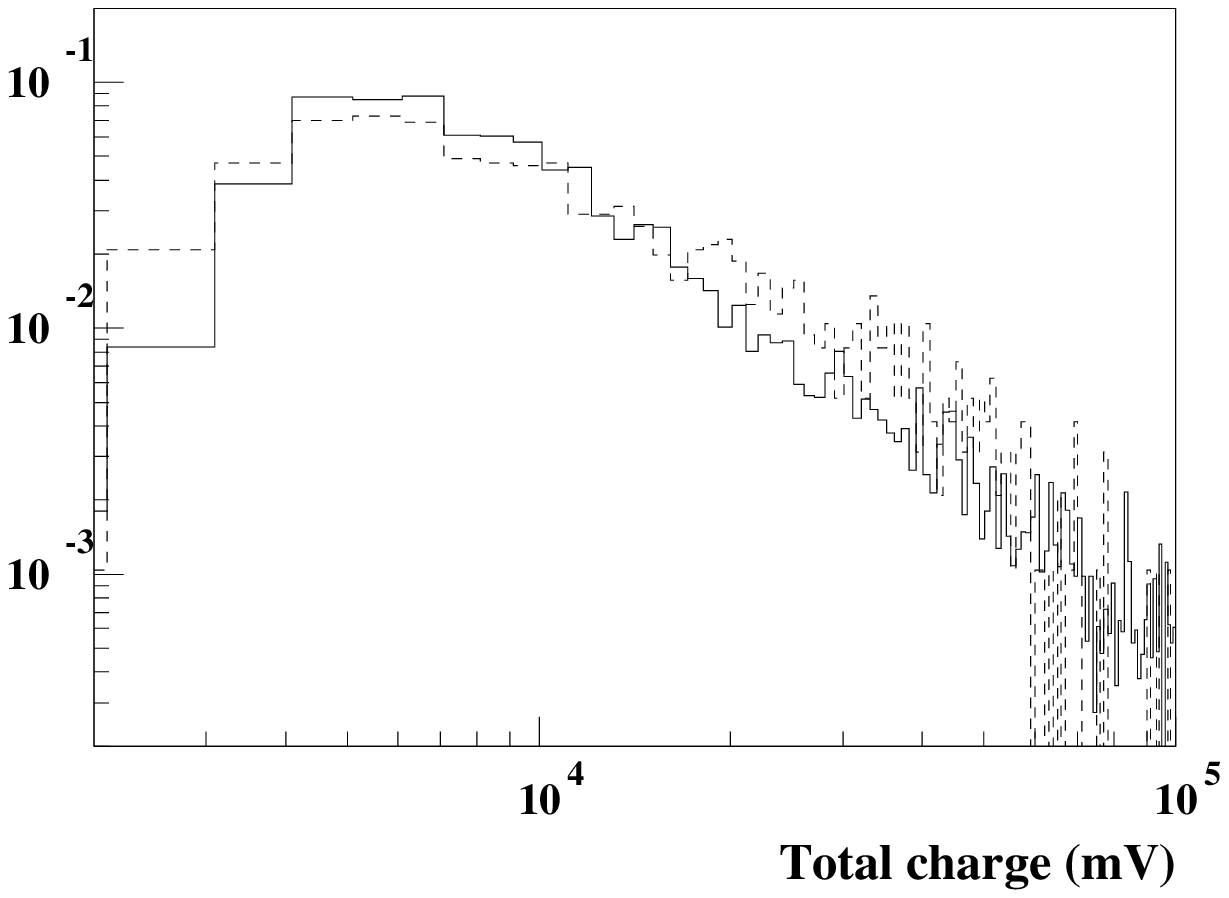,width=12cm,height=7cm,clip=,angle=0}
\caption{\it The number of showers as a function of ``total integrated charge''
in all 4 Cones in one shower. 
The dashed line are experimental data, the full
line is from the MC simulation. The curves were normalized for the
same number of showers.
The x-axis is in units of summed flash-ADC amplitudes in mV.
}
\label{totchar}
\end{figure}

Fig.\ref{totchar} displays the ``total charge'' spectrum both
in data and Monte Carlo. The ``total charge'' is determined
by integrating the area under all peaks detected 
in the flash-ADC traces adding
all four cones in one event.
Far above threshold the experimental spectrum follows a power law with a 
differential index of about
-1.6---which is much larger than that of the primary spectrum of -2.7.
The reason for this is a very large scatter in the 
correlation of total charge and energy.  
The Monte Carlo simulated spectrum looks qualitatively similar 
to the experimental data
but follows a slightly steeper index of about -1.9. 
One reason for this is that
far above the threshold the cutoff in simulated proton energy at 
10 TeV is already expected to
have a steepening effect on the MC spectrum.

\subsection{Effects of the small field-of-view on reconstructed shower
properties}

To gain the advantage of using many large mirrors with
only one central detector 
heliostats need to have a focal length about a factor 20 - 30 larger than
those of
the telescopes used for the imaging of VHE $\gamma$-ray showers.
For space reasons in the central tower 
the light detector at the focus cannot be scaled up by such
enormous factors. Moreover the construction of an imaging camera
for each heliostat would be prohibitively expensive.
These two factors force a crucial compromise in Cherenkov
detectors using heliostat fields: the field-of-view has to be
chosen about one to two orders of magnitude smaller
in solid-angle than in traditional Cherenkov telescopes.
\\
Our Monte Carlo simulations show that about 60$\%$ of the Cherenkov 
light of showers induced by gamma-rays with
small energies (100 GeV) is collected in the GRAAL setup, a number
that is acceptable when taking into account the large mirror area.  
Nevertheless, we find that the angular restriction has several 
disadvantageous effects. 

\subsubsection{Structure of timing front of proton induced showers}

\begin{figure}[ht]
\epsfig{
file=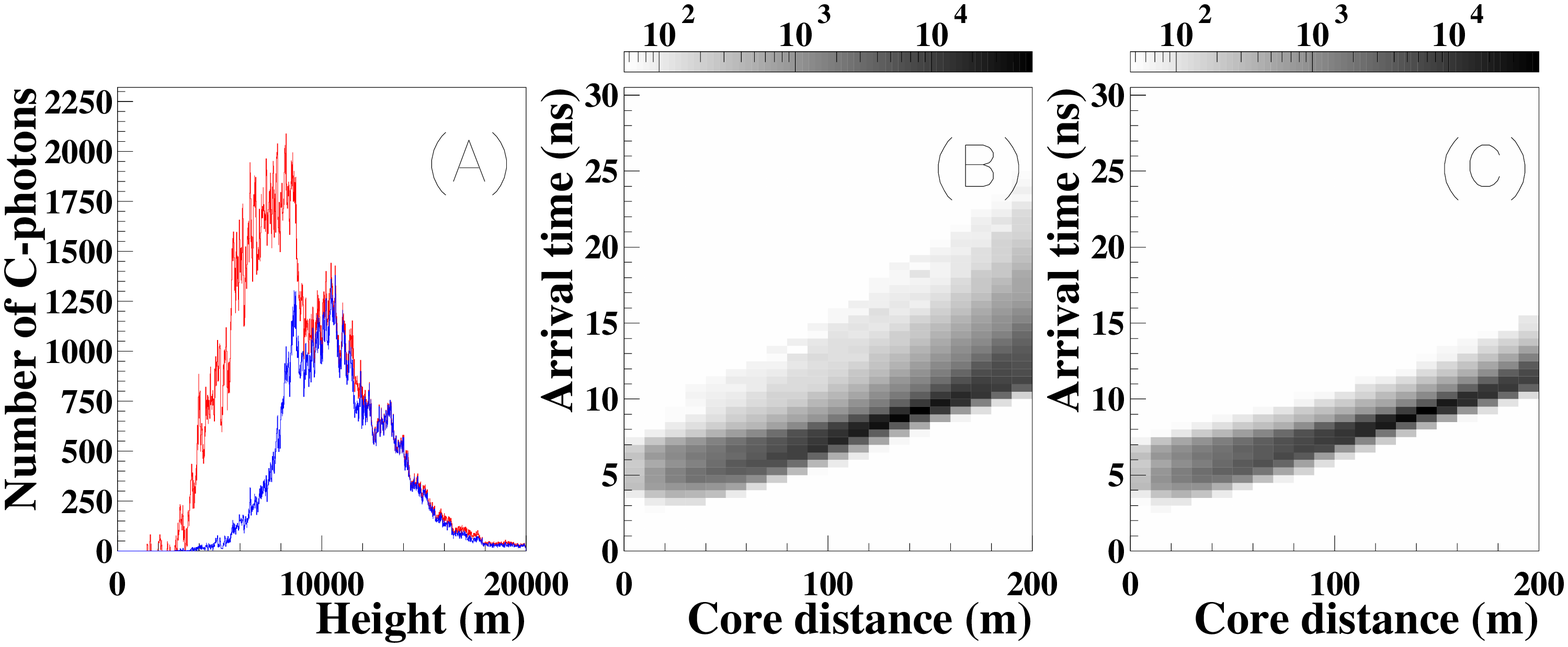,width=12cm,height=5cm,clip=,angle=0}
\caption{\it Time structure of a typical gamma-ray initiated shower.
b. The arrival time as a function distance from the core in meters
for a typical gamma shower. The shading
is proportional to the Cherenkov-photon density.
c. Same as b.
but only those photons with an arrival direction within 0.3$^{\circ}$
from the direction towards the shower maximum from a position on the ground are
displayed for the same showers.
a. Number of Cherenkov-photon emitting electrons
in the shower as a function of height a.s.l.
}
\label{tmcg}
\end{figure}

\begin{figure}[ht]
\epsfig{ 
file=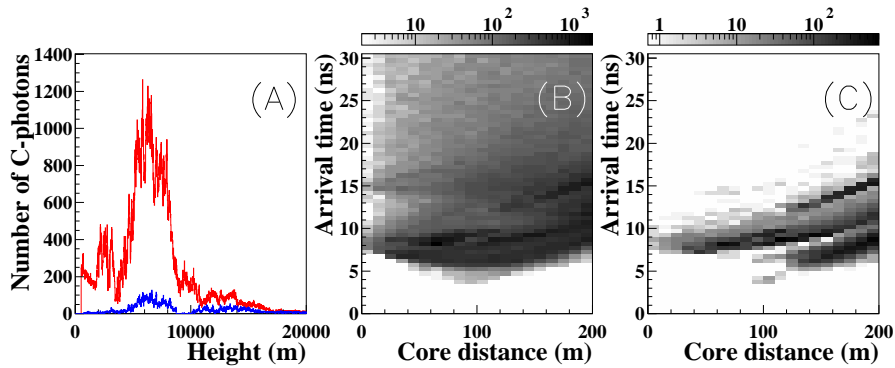,width=12cm,height=5cm,clip=,angle=0}
\caption{\it Time structure of a typical proton initiated shower.
The panels show the same quantities as in the previous fig.\ref{tmcg}.
Note that the proton emits 
a much smaller fraction of light within the restricted
field-of-view because of its larger angular extension.
}
\label{tmcp}
\end{figure}

\begin{figure}[ht]
\epsfig{
file=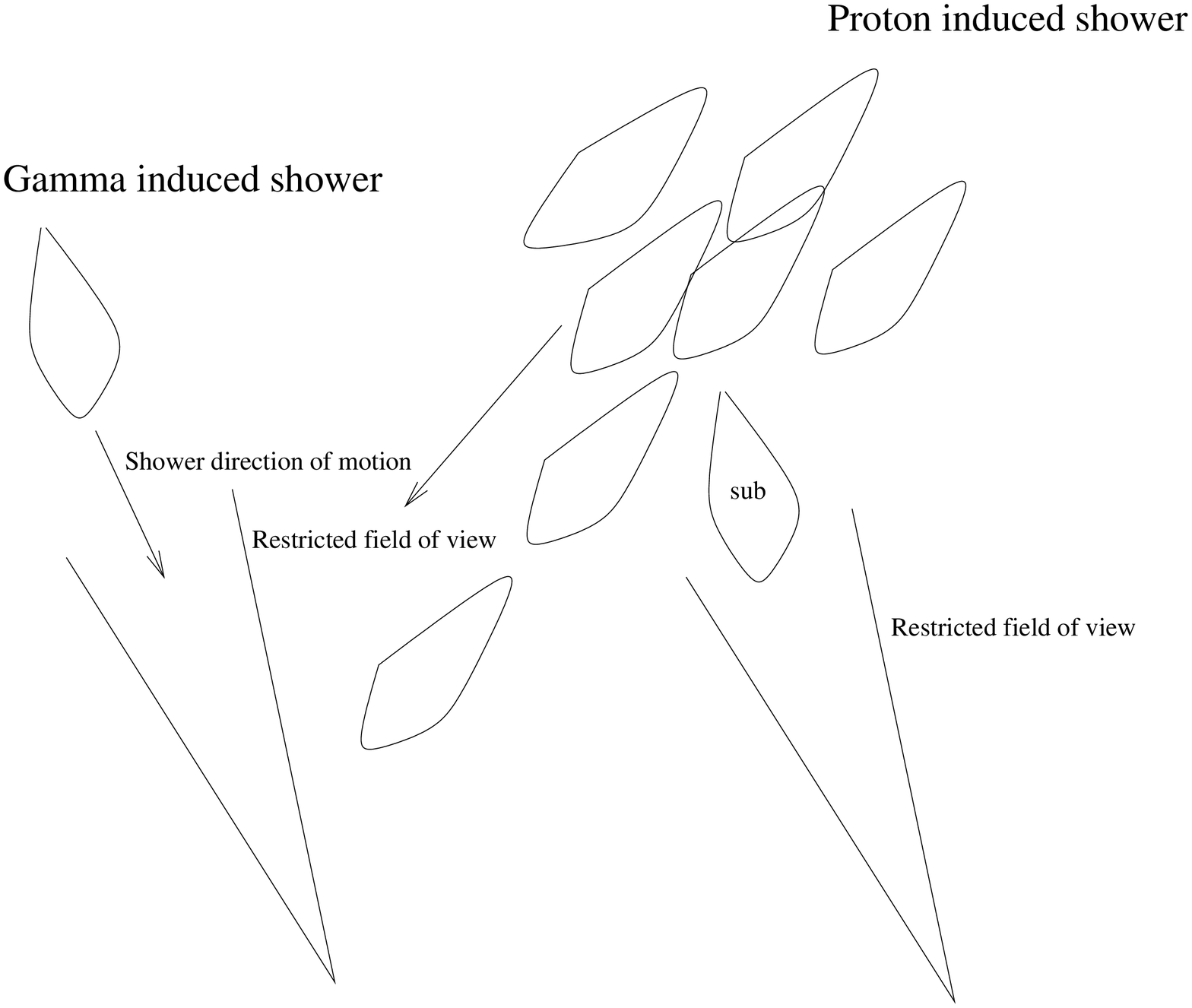,width=12cm,height=7cm,clip=,angle=0}
\caption{\it Sketch to illustrate the effect of a small-field
of view - necessitated by the heliostat-field approach -
on the determination of the timing structure.
A gamma-ray induced shower is symbolized in the left part of
the figure and a proton induced one with a slightly different
incidence direction on the right. The proton shower
is spatially more extended and symbolized as a collection
of small sub showers.
The restricted field of
view ``projects'' out sub showers in the central part of the shower
out of the more extended proton shower. Other more penetrating
and laterally extended sub showers - that increase the fluctuation
in the timing front -
do not contribute to the
light detected within the restricted field of view.
One sub shower with an incidence direction biased towards the 
pointing direction (symbolized by the label ``sub'') is preferentially
detected and thus biases reconstructed directions towards the pointing
direction.  
}
\label{geos}
\end{figure}

It is well known that the arrival times in proton-induced showers have a much 
wider scatter around the mean arrival time than in gamma-induced showers due
to their more irregular development in the atmosphere\cite{gammahad}.
The experimental determination of this scatter has been proposed to be an 
efficient
method for gamma/hadron separation\cite{indiant}.
Fig.\ref{tmcg}b. shows the structure of the shower front of a typical 
gamma, fig.\ref{tmcp}b. of a typical proton shower
from the Monte Carlo simulation without simulation of the detector.
The larger scatter of the proton shower is evident.
In panel c. of these figures the shower front is 
shown with a restriction
on the incident angle of the photon. 
Only photons with an incident angle different
by less than 0.3$^{\circ}$ from the direction pointing 
towards the shower maximum
from a position on ground were retained. 
This restriction has a very
similar effect to the small field of view dicussed above.
Upon angular restriction the shower front narrows both for protons and gammas,
but the effect is stronger for the protons. 
Panels a. of these figures demonstrate the cause
of this behaviour. The restricted field of view mainly prevents the detection
of Cherenkov photons emitted far below the maximum at about 11 km height.
Deviations from the ideal spherical timing-front are 
mainly due to the deeply penetrating
part of the shower. 
Protons are more penetrating and are therefore more affected
by the angular restriction.
\\
The total effect is that with a small field of view, protons and gammas have 
virtually identical, nearly spherical shower fronts, 
with very little scatter,
displayed and compared with experimental data in fig.\ref{timed}. 
This makes efficient gamma/hadron
separation with timing methods in heliostat fields all but impossible (see 
figs.\ref{timed},\ref{geos}).

\subsubsection{Reconstructed direction of proton induced showers}

Another important method to discriminate gamma- and proton induced showers
is to exclude all showers that do not arrive from the source
direction within the angular resolution as determined with a fit to the
timing front. With a restricted field of view there is a bias
of the shower direction reconstructed from timing information
towards the source direction (see fig.\ref{geos}).
The field-of-view ``selects'' a part of the shower
which lies towards the shower maximum of a shower arriving from the source direction.
The timing-fit then finds the direction of this subpart of the shower, which
is biased towards the source direction. In Monte-Carlo simulations of proton
induced showers we
found that the mean difference between true shower direction and reconstructed
shower direction is 0.71$^{\circ}$ $\pm$ 0.002$^{\circ}$ (statistical
error), whereas 
the mean difference between source 
direction and reconstructed shower 
direction is only 0.44$^{\circ}$ $\pm$ 0.002$^{\circ}$ (statistical error).
This bias decreases the fraction of proton showers which can be excluded
due to their angular distance to the source direction.

\subsubsection{Energy resolution}
The restriction in the field of view decreases the energy resolution
progressively for large showers, because the fraction of the
shower image seen cannot be inferred.
We find that near
our energy threshold for gamma rays
the resolution derived from choosing the total charge recorded in
all 4 cones as simple primary-energy estimator is about 110$\%$ and
worsens rapidly for higher energies.

\section{Data selection}
\label{datasel}
\subsection{Detector condition}
\label{detsel}
Only nights in which all four detector channels
and the heliostats in the field were functioning normally according to the 
recorded monitor files 
were chosen for further analysis.
\subsection{Meteorological selection}
\label{metsel}
It was found that the reconstruction quality depends
on the atmospheric transmission. For example in nights which were visibly
hazy with a high relative humidity above 80$\%$ (a relatively frequent
nightly weather condition at the PSA),
the total trigger rate
was low, the ratio of well reconstructed events
to events with a misreconstructed angular direction
(called ``PT'' below) was reduced by up to a factor 2 
and the ${\rm lsq_t^2}$ 
of the fit to the timing front significantly increased.
This is probably the result of selective absorption, by which Cherenkov light 
from the deeply
penetrating part of the airshower, with increased temporal fluctuations,
dominates the recorded signal.
As $\gamma$-induced showers develop mainly in the upper atmosphere
a selection of data without selective absorption is important.
Besides a relative humidity below 70$\%$ and generally clear skies
we required the following criteria from the reconstructed data
of a given night. The parameter
limits for each individual pointing direction
were chosen such that a set of ``good'' nights
---defined as showing fairly constant parameter values---was retained.
The parameter limits thus slightly varied for different pointing
directions. 
\\
First a cut to exclude unstable weather conditions was applied. For this
the fraction of events with a reconstructed angle 
far from the source direction was chosen.
\\
{\bf Condition 1: 0.95 $<$ RO $<$ 1.05}
\\
RO = (Number of events with reconstructed direction $>$ 3$^{\circ}$ from
pointing direction ON source) / 
(Number of events with reconstructed direction $>$ 3$^{\circ}$ from
pointing direction OFF source)
\\
The other two run-cut criteria are meant to exclude nights with 
low atmospheric transmission.
\\
{\bf Condition 2: Rate after all software cuts in OFF source direction $>$ 50/min}
\\
{\bf Condition 3: PT $>$ 0.8}
\\
PT = (Number of events with reconstructed directions $<$ 1$^{\circ}$ from pointing 
direction OFF source)/
(Number of events with reconstructed directions $>$ 3$^{\circ}$ from pointing direction OFF source)
\\
These ``meteorological cuts'' are severe under the weather conditions
at the PSA. In the data sample on Crab in February/March 2000 
only 22$\%$ of all data 
taken on the Crab pulsar passed all cuts.

\section{Data reduction---the problem of different conditions in the
source and OFF-source region}
\label{datared}
A fundamental problem of all Cherenkov experiments---specially 
for those attempting to detect an excess due to gamma-rays in the
total rate---is the fact that the night-sky background between 
ON- and OFF-source
differs in general. 
This can influence the counting rate and analysis efficiency
in various ways. This problem is most critical for the heliostat-array based
experiments because they aim to work with a trigger threshold 
near to random fluctuations of the night-sky background in a single
channel.
We discuss the observed differences in the ON- and OFF-source
region in section \ref{offon}. Sections \ref{hardp} and
\ref{softp} discuss the effect of a difference in the ON- and OFF-source
intensity of the NSB on the total rate and the reconstruction,
respectively. Section \ref{excessc} describes the method we finally
chose to calculate an excess of events in the ON-source
direction.

\subsection{Detailed comparison of conditions ON and OFF source}
\label{offon}
For the counting conditions chosen in the 1999/2000 season,
the ``q-trigger'' (see section \ref{trigger}) leads to random event
triggers due to night-sky noise. Because this random rate
is very sensitive to the night-sky background, slightly
higher NSB levels in OFF (as observed for all potential sources
see tables (\ref{crabresult} - \ref{pseudoresult})) 
produce a higher event-trigger rate in OFF
(see entry ``raw events'' in tables \ref{crabresult}-\ref{3egresult})
The random rate was calculated
from the recorded single rates. A discussion
of the total rate after a correction for the
random trigger and other small effects is given below
in section \ref{hardp}.
From test data it
was shown that the reconstruction of the timing shower
front always fails for random-trigger events, so that
the event number ``after reconstruction'' (``rec. events'' in
tables \ref{crabresult} - \ref{3egresult}) 
is expected to be free from night-sky background induced random
triggers.
\\
All 4 sources discussed in this paper show a slightly
higher NSB in the OFF-source region. This effect is
most pronounced for the source 3C 454, where---from the data
reported in table \ref{3c454result}---the current (q-rate) is 
13 (25)$\%$ higher in OFF than in ON.
However, the measured RMS fluctuation
is only 0.4$\%$ higher in OFF than in ON and this difference
is smaller for the other sources (0.04$\%$ for the Crab pulsar).
By measuring the random noise in complete darkness, we determined
a constant night-sky background independent noise level with a RMS
of 0.8658. Subtracting this constant noise quadratically 
from the total noise we get the contribution from the NSB alone
(number in brackets in third column of table \ref{crabresult} - 
\ref{pseudoresult}). For the source with the largest difference in noise
level, the NSB-induced component differs in ON- and OFF-source position
by about 2.5$\%$, so that the difference in brightness at the
two positions can be estimated to be about 5$\%$.
\\
An effect that is very difficult to remove is a slight expected reduction
in the trigger threshold due to a higher NSB. 
Due to fluctuations, smaller events can cross the trigger threshold.
The opposite effect---that large events are decreased in amplitude
due to fluctuations and fail to cross the threshold---happens less often 
due to a CR spectrum that steeply
falls with amplitude.
\begin{table}[ht]
\vspace{-10pt}
\caption{\it {\bf Current} (mean of 4 Cones), {\bf q-rate}: single trigger rate of charge 
integrating channel (mean of Cone 1+2), {$\bf \sigma_{\rm NSB}$}:
RMS fluctuation of the measured
NSB (in flash-ADC units) in the first 100 channels (before signal),
number in brackets is NSB induced background alone, {\bf log(mean q)}:
base-10 logarithm of
mean net-charge (in flash-ADC units) of all events in sample,
{\bf raw events}: all hardware-triggered events which traces
were recorded,
{\bf rec. events}:
number of events after angular reconstruction and software trigger,
{\bf centr. events}: normalized number of events in central
angular region (within 0.7 degrees of pointing direction), calculated
as explained in section \ref{excessc}.
Rows are for the samples with pointing towards
the Crab pulsar (``ON'') and on a sky position (``OFF'')
with a right ascension 2.625 degrees
larger than in the ON direction.
The total data-taking time ON was 430 minutes with an equal amount of OFF time.}
\label{crabresult}
\begin{center}
\begin{tabular}{lcccc}
  &  current [$\mu$A] & q-rate[kHz] & $\sigma_{\rm NSB}$ [ADC units] &  log(mean q)   \\
\hline
\hline
ON  & 19.0 $\pm$ 0.4 & 1.35 & 0.9493 (0.3893) & 2.940 $\pm$ 0.004   \\ 
OFF & 19.3 $\pm$ 0.3 & 1.49 & 0.9497 (0.3902) & 2.937 $\pm$ 0.004   \\ 
\hline
EXCESS & -0.3 & -0.14 & -0.0004 (-0.0009) & 0.003 $\pm$ 0.006  \\ 
\hline
\end{tabular}
\begin{tabular}{lccc}
 & raw events &  rec. events &  centr. events   \\
\hline
\hline
ON  & 68702 & 33384  & 9415  \\ 
OFF & 75198 & 33056  & 8678  \\ 
\hline
EXCESS & -6496 $\pm$ 379 & 328 $\pm$ 258 & 737 $\pm$ 165 \\ 
\hline
\end{tabular}
\end{center}
\end{table}

\begin{table}[ht]
\vspace{-10pt}
\caption{\it Entries as in table \ref{crabresult}
for the samples with pointing towards
the radio source 3C 454.3 (``ON'') and on a sky position (``OFF'')
with a right ascension 2.625 degrees
larger than in the ON direction.
The total data-taking time ON was 550 minutes with an equal amount of OFF time.}
\label{3c454result}
\begin{center}
\begin{tabular}{lcccc}
  &  current [$\mu$A] & q-rate[kHz] & $\sigma_{\rm NSB}$ [ADC units] &  log(mean q)   \\
\hline
\hline
ON  & 17.7 $\pm$ 0.4 & 3.1 & 0.9505 (0.3922) & 3.119 $\pm$ 0.003   \\ 
OFF & 20.3 $\pm$ 0.3 & 4.1 & 0.9540 (0.4006) & 3.113 $\pm$ 0.003   \\ 
\hline
EXCESS & -2.6 & -1.0 & -0.0035 (-0.0084) & 0.006 $\pm$ 0.004  \\ 
\hline
\end{tabular}
\begin{tabular}{lccc}
 & raw events &  rec. events &  centr. events   \\
\hline
\hline
ON  & 42516 & 30570  & 7525  \\ 
OFF & 44949 & 30889  & 7625  \\ 
\hline
EXCESS & -2433 $\pm$ 296 & -319 $\pm$ 248 & 54 $\pm$ 141 \\ 
\hline
\end{tabular}
\end{center}
\end{table}

\begin{table}[ht]
\vspace{-10pt}
\caption{\it Entries as in table \ref{crabresult}
for the samples with pointing towards
the unidentified $\gamma$-ray
source 3EG J1835+59 (``ON'') and on a sky position with a 
right ascension 2.625 degrees
larger than in the ON direction.
The total data-taking time ON was 490 minutes with an equal amount of OFF time.
The result for the central region is not given for this source because
the quality of the angular reconstruction was strongly decreased for its
direction pointing towards the north.}
\label{3egresult}
\begin{center}
\begin{tabular}{lcccc}
  &  current [$\mu$A] & q-rate[kHz] & $\sigma_{\rm NSB}$ [ADC units] &  log(mean q)   \\
\hline
\hline
ON  & 15.5 $\pm$ 0.6 & 1.7 & 0.9528 (0.3977) & 3.122 $\pm$ 0.002   \\ 
OFF & 15.8 $\pm$ 0.6 & 1.8 & 0.9519 (0.3956) & 3.116 $\pm$ 0.002   \\ 
\hline
EXCESS & -0.3 & -0.1 & 0.0009 (0.0021) & 0.006 $\pm$ 0.003  \\ 
\hline
\end{tabular}
\begin{tabular}{lccc}
 & raw events &  rec. events &  centr. events   \\
\hline
\hline
ON  & 45984 & 21639  & -  \\ 
OFF & 46431 & 21772  & -  \\ 
\hline
EXCESS & -447 $\pm$ 304 & -25  $\pm$ 212 & - \\ 
\hline
\end{tabular}
\end{center}
\end{table}

\begin{table}[ht]
\vspace{-10pt}
\caption{\it Entries as in table \ref{crabresult}
for the samples with pointing towards
a ``pseudo source'' at right ascension = 330.68 degrees and 
declination = 40.28 degrees (``ON'')
and on a sky position with a right ascension 2.625 degrees
larger than in the ON direction.
The total data-taking time ON was 
250 minutes with an equal amount of OFF time.}
\label{pseudoresult}
\begin{center}
\begin{tabular}{lcccc}
  &  current [$\mu$A] & q-rate[kHz] & $\sigma_{\rm NSB}$ [ADC units] &  log(mean q)   \\
\hline
\hline
ON  & 16.6 $\pm$ 0.5 & 4.3 & 0.9564 (0.4063) & 2.991 $\pm$ 0.003   \\ 
OFF & 17.3 $\pm$ 0.5 & 5.7 & 0.9588 (0.4119) & 2.993 $\pm$ 0.003   \\ 
\hline
EXCESS & -0.7 & -1.4 & -0.0024 (-0.0056) & -0.002 $\pm$ 0.005  \\ 
\hline
\end{tabular}
\begin{tabular}{lccc}
 & raw events &  rec. events &  centr. events   \\
\hline
\hline
ON  & 24119 & 13136  & 2295  \\ 
OFF & 26911 & 13272  & 2299  \\ 
\hline
EXCESS & -2792 $\pm$ 226 & -136 $\pm$ 136 & -7 $\pm$ 76 \\ 
\hline
\end{tabular}
\end{center}
\end{table}

This effect was studied by calculating the total mean charge of all ON versus 
all OFF source events (
see tables \ref{crabresult} - \ref{pseudoresult} 
entry ``mean q''). If small events are preferred,
the total mean charge should be smaller by a certain factor f$_b$
with increased NSB.
The total rate should be increased by a factor of roughly f$_{b}^{1.4}$
for our setup.
\\
It can be seen from the results in the tables 
(\ref{crabresult} - \ref{3egresult}) that within the
statistical error of typically somewhat less than a percent
the total mean charge is the same for all
sources. However, within this error a significant reduction of
the threshold - one  which would produce a reduction in ON-OFF rates
of the same order of magnitude as an expected signal from the
crab nebula - cannot be excluded in this way.

\subsection{Effect of NSB differences on total trigger rate -
Monte-Carlo simulation}
\label{hardp}
\begin{figure}[ht]
\epsfig{
file=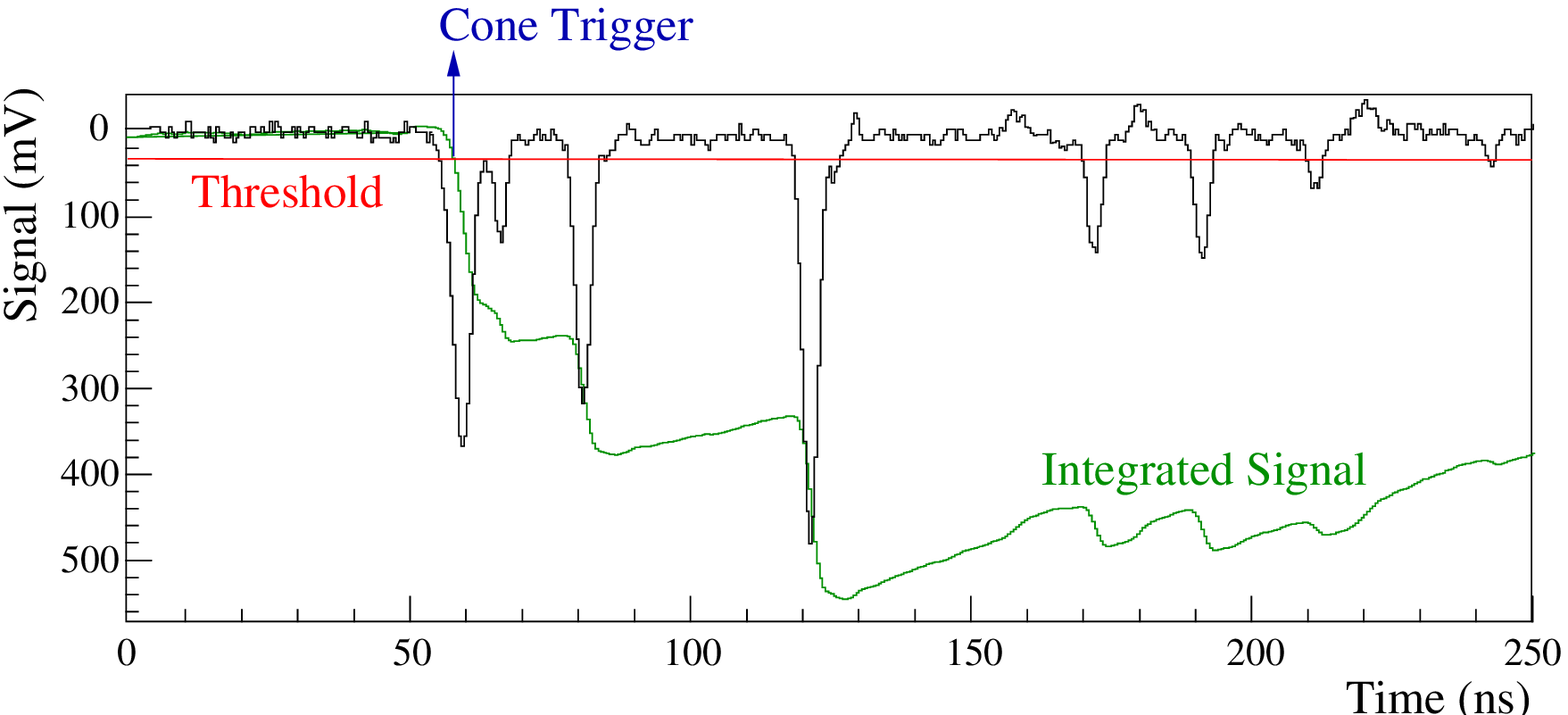,width=12cm,height=7cm,clip=,angle=0}
\caption{\it 
Trigger simulation in the detector Monte Carlo.
A trace with high time resolution is integrated in the
electronic chain of the ``charge(q)-trigger''(section \ref{trigger}),
resulting in the signal labelled ``Integrated signal''. Once
this signal surpasses the threshold level, a Cone Trigger
is initiated. 
}
\label{ctrigger}
\end{figure}

The effect of the NSB differences on the total trigger rate was
simulated by raising the amount of random noise by 5$\%$ over its
usual value. The detector Monte Carlo models the electronic
pulse shaping and
the response of the discriminator in detail (see fig.\ref{ctrigger}),
and so the effective change in threshold, due  to the increased
noise level could be deduced to be about 6$\pm$ 2$\%$.
Extrapolating, we deduce an expected spurious excess at the OFF
source position for the source with the largest difference in 
ON and OFF noise (3C454, see table \ref{3c454result}) of about 1$\%$, this
corresponds to about 1.4 $\sigma_{\rm stat}$ in this case.
As the difference in the noise levels between ON and OFF is
smaller in the case of the other 3 sources discussed in this paper,
this effect does not yet contribute significantly.
However, it is clear that a very careful correction for
it becomes necessary when the available statistics grows.

\subsection{Effect of NSB differences on reconstruction - Software padding}
\label{softp}
\begin{figure}[ht]
\epsfig{
file=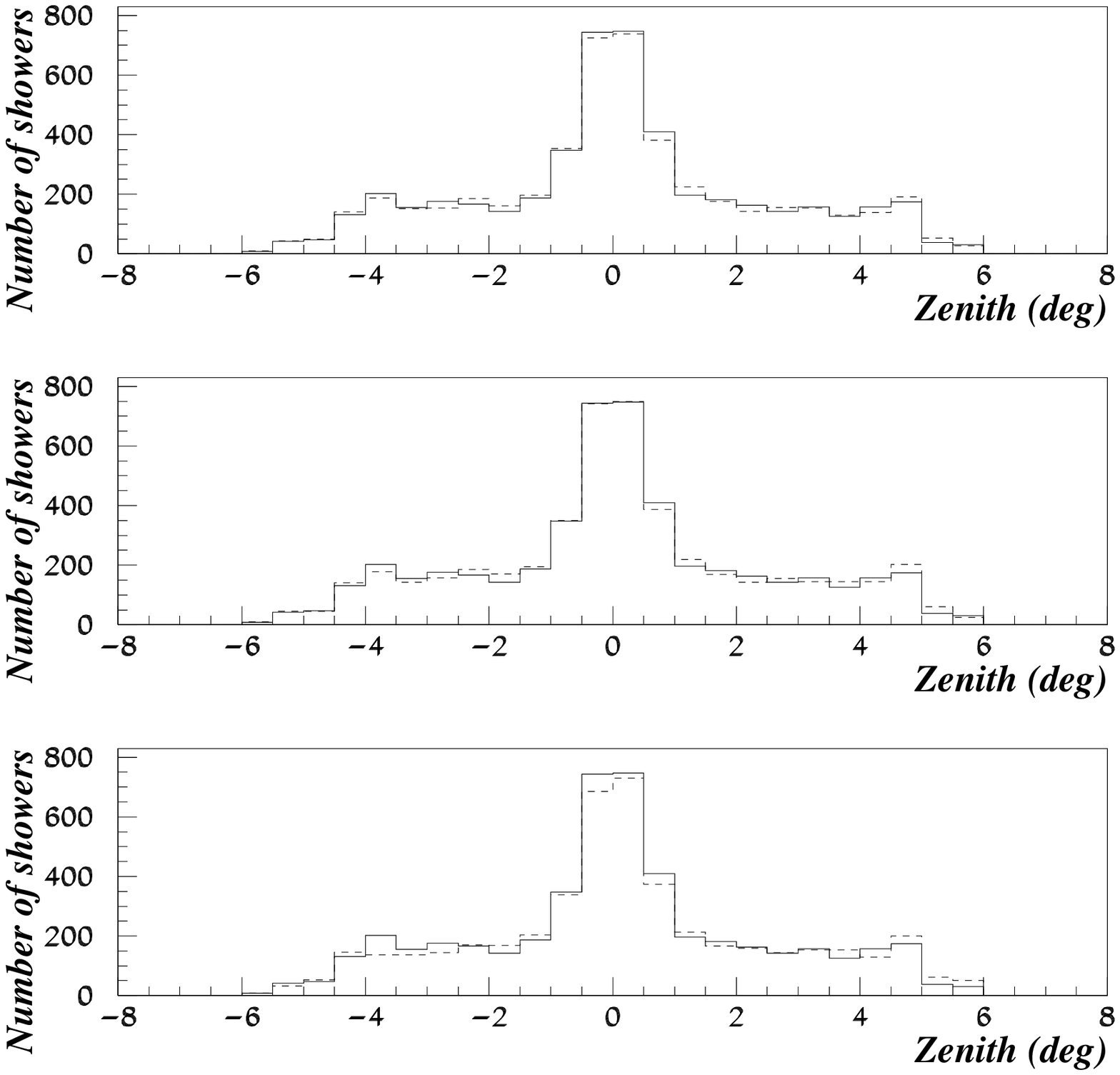,width=12cm,height=7cm,clip=,angle=0}
\caption{\it The number of showers as a function of difference in zenith
angle to the source direction reconstructed with the experimental NSB
(full line) and NSB increased on the software level (dashed line) by
0.06$\%$ (upper plot), 0.9$\%$ (middle plot) and 30$\%$ (lower plot). In the
lower plot a decrease of the fraction of events within the central
region is obvious.
}
\label{softpad}
\end{figure}

Finally a difference in NSB leads to a slightly different noise levels
in ON and OFF data. For example for the Crab data the RMS noise in the
ON-source data was found to be about 0.5$\%$ smaller than in the OFF-source data.
The effect of this difference on the reconstruction procedure
was studied by artificially adding noise at the software level 
(``software padding''). Fig.\ref{softpad} demonstrates that 
the fraction of events near the source direction
(``PT'' of section \ref{metsel})
decreases with increasing NSB, but that the effect is important
only at relatively large increases on the order of a few percent.
From the results shown in figure \ref{softpad}
it was derived that an increase of RMS noise by 1$\%$
decreases the overall reconstruction efficiency by about 0.4$\%$ and
the peak to tail ratio PT (section \ref{metsel}) by 0.8$\%$. 
This effect remains small for the observed 
fractional differences of the RMS NSB-noise in 
ON and OFF (on the order a few tenths of a percent 
at maximum (see table \ref{crabresult}-
\ref{pseudoresult})),
and was neglected in the present analysis.
It must be noted that software padding is not a perfect simulation
of the real conditions, because the influence of the NSB on the
hardware-trigger condition---which can in principle
influence shower properties and reconstruction---is not simulated.

\subsection{Calculation of the excess}
\label{excessc}
To avoid the problem mentioned in section \ref{hardp} we chose
a method that normalizes any excess to the ratio of ON- and OFF-source
events for the final results reported in the next section.
\\
The normalized excess EXCESS$_{\rm{n}}$
was calculated according to the following equation:
\begin{equation}
\rm{EXCESS_n = ON_{in} - OFF_{in} \left(ON \over OFF\right)_{out}} 
\label{exc}
\end{equation}
Here $\rm{(ON,OFF)_{in}}$ stands for the number of events within 0.7$^{\circ}$
from the source and OFF-source  
direction, respectively whereas $\rm{(ON,OFF)_{out}}$ stands for the number of 
events with directions
deviating more than 2$^{\circ}$ from the source direction.
The statistical error of EXCESS$_{\rm n}$, ERR$_{\rm n}$ 
was calculated according to:
\begin{eqnarray}
\rm{ERR_n}= \left( ON_{in} + OFF_{in} \times 
\left( ON \over OFF \right)_{out}^2 +
\left[{(1 + \left(ON \over OFF\right)_{out}) ON_{out} \times OFF_{in}^2} \over
OFF_{out}^2 \right] \right)^{0.5}
\label{err}
\end{eqnarray}

\section{Results}
\label{results}
\subsection{Crab pulsar}
\begin{figure}[ht]
\epsfig{
file=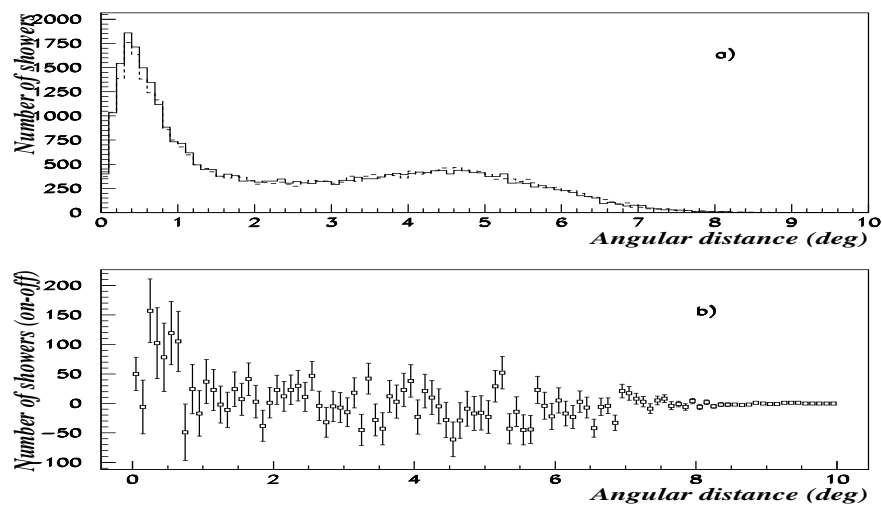,width=12cm,height=7cm,clip=,angle=0}
\caption{\it  The upper plot (a.) shows the
number of events as a function of angular distance of 
reconstructed direction from source direction for ON-source events
(full line) and OFF-source events (dashed line).
No normalization
of any kind was applied to this plot.
The lower plot (b.) shows the difference ON - OFF, normalized
to the number of events in the outer angular region, according to
eq.\ref{exc}
Data from the Crab pulsar
taken under good meteorological conditions according to the
cuts discussed in section \ref{metsel}
were used. The statistical errors of the individual bins are
shown.}
\label{evangcrab}
\end{figure}

Several parameters of the data set taken on Crab pulsar are
presented in table \ref{crabresult}. A parameter
n$_t$=5 for the software-trigger was chosen (section \ref{evreco_tri}).
Fig. \ref{evangcrab} shows the number of events as function
of angular distance from the source direction, both for 
ON- and OFF-source
direction and the normalized difference ON-OFF. An 
excess of events in the angular region expected from Monte-Carlo 
simulations
(fig.\ref{evangcrab}) is seen, we find
\\
EXCESS$_n$ = 737 $\pm$ 165 
\\
calculated according to eqns (\ref{exc},\ref{err}).
This corresponds to a 4.5 $\sigma$ excess and a mean excess rate
EXCESS$_{\rm nr}$ = 1.7/min.
Fig.\ref{exczenaz} displays the excess as a projection onto
zenith and azimuth axis respectively.
An integral flux $\phi_{int}$ is calculated from this excess 
according to:
\begin{equation}
{\rm \phi_{int} = (EXCESS_{nr}/r_{\gamma}) 
(r_p/r_{\rm obs}) t_c \phi_{whipple}}
\label{flux}
\end{equation}
Here $\phi_{\rm whipple}$=
$\int_{\rm E_{\rm thresh}}^{\infty}$ 
3.3 $\times$ 10$^{-7}$ E$^{-2.4}$ m$^{-2}$ sec$^{-1}$ TeV$^{-1}$ dE
is the integral gamma-ray flux from the Crab above a threshold energy 
E$_{\rm thresh}$
as observed by the Whipple collaboration\cite{whipple}. 
r$_{\gamma}$ is the 
gamma-ray rate expected in GRAAL from the Monte-Carlo simulated
effective area for gammas of fig.\ref{detarea}
based on this flux (0.011 Hz).
Note that the absolute Whipple flux cancels in eq.\ref{flux}, and we
only adopt the spectral index from ref.\cite{whipple}. 
r$_p$ is the proton rate expected in GRAAL
on the basis of the known proton flux $\phi_{ref}$ and the effective 
area for protons of fig.\ref{detarea} (4.0 Hz). r$_{obs}$ is
the observed cosmic-ray rate in the final reconstructed sample, corrected
for dead time (1.6 Hz).   
The factor (r$_{\rm obs}$/r$_p$) is an
empirical correction for the fact that
our Monte-Carlo calculated 
proton effective area fig.\ref{detarea} 
predicts a somewhat higher proton rate
than observed. 
t$_c$ is a correction factor for the fact that some photons are
expected in the ``outer angular region'' and was determined as
2.2 (1.4) from weighted (unweighted) Monte Carlo data.
The weighted value was chosen for the final result.
The final integral flux above threshold
assuming a differential spectral source index of -2.4 is:
\\
$\phi_{int}$ = 2.2 $\pm$ 0.4 (stat) $^{+1.7}_{-1.3}$ (syst) $\times$ 
10$^{-9}$ cm$^{-2}$ sec$^{-1}$ above threshold
\\
The systematic error of our flux determination
is dominated by the uncertainty in absolute 
light-calibration.
The conversion factor ``total light at entrance of Cone vs. flash ADC channel''
used in the Monte-Carlo simulation for the effective detection area
can be compared with the one derived from  
the LED based method described in section \ref{cal} 
for all four Cones.
The relative difference:
\\
((predicted ADC channel Monte Carlo) - (predicted ADC channel LED))/
(predicted ADC channel Monte Carlo)
\\
was 21$\%$,-31$\%$,-13$\%$,+29$\%$ for Cone 1-4. From this we 
estimate a systematic error
of 30$\%$ for this conversion. We estimate a similar error
due to uncertainties in the Monte Carlo simulations
between the primary and the entrance of Cones which increase 
the error in absolute
light calibration to about 42$\%$, corresponding to a flux error of 
about $_{-60}^{+81} \%$.
Another important source of overall systematic error is 
the systematic errors of
t$_{\rm c}$ (35$\%$) in which uncertainties 
in the spectral weighting procedure and
the detailed simulation of the trigger enter and which
was added in quadrature. The final adopted systematic error
is  $_{-69}^{+88}\%$.
This result is compared with other flux determinations 
in fig.\ref{final}.

\subsection{Other potential sources}
\begin{figure}[ht]
\epsfig{
file=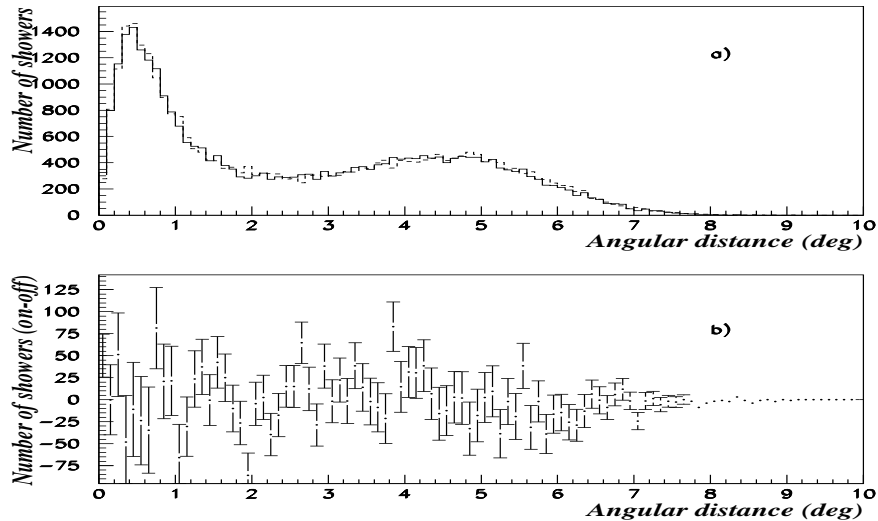,width=12cm,height=7cm,clip=,angle=0}
\caption{\it Good-weather data of the potential source 3C 454.3
plotted as in fig.
\ref{evangcrab} 
}
\label{evang3c454}
\end{figure}

\begin{figure}[ht]
\epsfig{
file=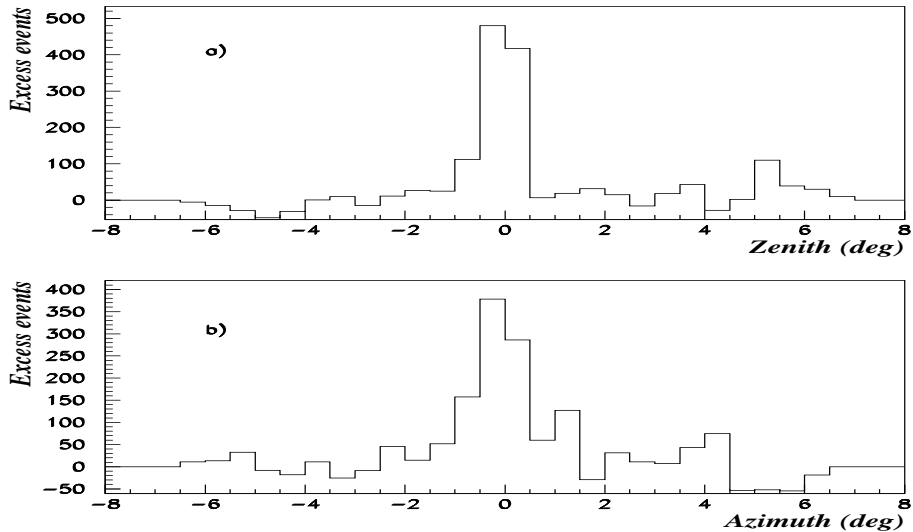,width=12cm,height=7cm,clip=,angle=0}
\caption{\it The difference of the number of 
events in ON source direction and OFF source direction
for the Crab data sample shown in fig.
\ref{evangcrab}
as a function of deviation of the
zenith (upper plot a.) and azimuth angle 
(lower plot b.) from the source direction.
}
\label{exczenaz}
\end{figure}

\begin{figure}[ht]
\epsfig{
file=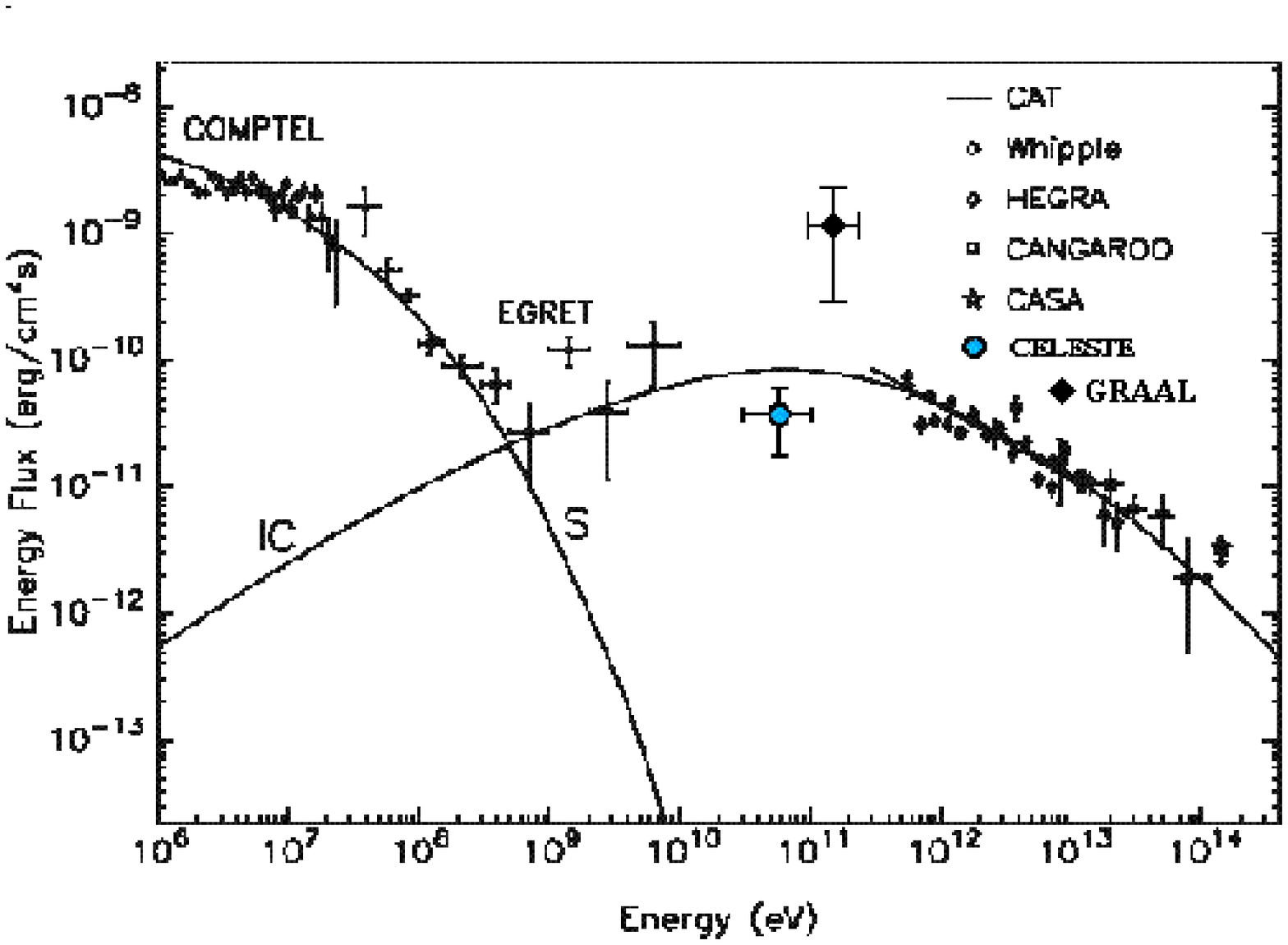,width=12cm,height=7cm,clip=,angle=0}
\caption{\it The energy flux as a function of energy 
as determined here (diamond) compared to
determinations in other experiments (adapted from \cite{celestecrab}).
}
\label{final}
\end{figure}

Fig. \ref{evang3c454} and table \ref{3c454result}
present a data set---of very similar size and taken under similar
conditions as the data on the Crab pulsar---on the potential gamma 
source 3C 454.3. n$_t$=7 was chosen.
No significant excess
can be seen. Table \ref{3egresult} and \ref{pseudoresult} present
the analysis on the potential source 3EG J1835+59 (n$_t$=9)
and a ``pseudo'' source (n$_t$=7),
with data taken with a pointing towards a dark spot in the night sky.
In both cases the results are in agreement with no gamma-ray emission
from these sources.
The former source lies towards the north of the heliostat field.
As discussed in section \ref{diff_cel} this leads to a worse
reconstruction of direction and a derivation of number
of ``central events'' does not make sense.

\subsection{Excess in total rate}
If the detected excess (discussed in section \ref{results}) is real, one
can estimate that there
should be an excess of $\approx$ 2270 events within our measuring
time. On the other hand, extrapolating the Whipple flux for
Crab nebula (\cite{whipple}) at our energy threshold, only 355 excess
events are expected.
\\
Due to
our trigger setup it can happen that fluctuations of
the NSB alone trigger events. 
The rate of these ``accidental trigger'' can be calculated from the single
rates (properly taking into account the trigger logic discussed
in section \ref{trigger}) and
subtracted from the total rate. For the sequence trigger the 
probability that cones 1 and 2 trigger at the same time accidentally
has been calculated from the
individual sequence trigger rates of each cone.
For the charge trigger
the accidental events are given by the probability of 3 cones out of 4
triggering simultaneously due to the individual q-rates at each
cone. The probability of accidental events is calculated every 2
seconds, so that peaks of high intensity (e.g. due to the light
of a car) can be detected.
The number of accidental events rises with the individual
sequence and q rates. With the new setting of season 2000/2001 
(data from this season are not discussed in this paper) the
individual rates have been lowered so that the total rate of real events
is still the same as for season 1999/2000 but there are no more accidental
events.  
Other corrections are related to the dead time of the setup.
During each run, calibrations are done regularly to verify time and
amplitude of the peaks and gain properties of the electronic chain
(see section \ref{cal}). The time used for the
calibrations can be slightly different in ON and OFF periods. It may
also happen that some time is lost due to the switching off of the
PMTs (for safety reasons, if the currents increase above 35$\mu$A
the PMTs switch off automatically for $\approx$ 15 seconds). These 2
factors can produce a difference in measuring time in ON and OFF
periods. A correction factor is applied so that periods ON and OFF have
exactly the same time of measurement. 

Table \ref{accident} right column shows the results of a careful
correction for these effects for the
same four data
samples as used for tables \ref{crabresult} - \ref{pseudoresult}
(see section \ref{datared}). In the last
column all the effects have 
been corrected. For the case of the
Crab nebula there is an excess in the OFF position
of 7234 events in the hardware-triggered events. After subtraction of
accidental events and corrections for dead time, the excess in the OFF
position
is only 443 events, which is within the statistical fluctuations. For
orientation, a difference in the energy threshold of cosmic-ray protons
between ON and OFF of only 5 GeV at an energy threshold of 2 TeV already
produces a difference of 550 events
for the same time of measurement and using the known cosmic-ray proton
flux and a constant effective area of 8000 m$^2$.
\begin{table}[t]
\caption{\it Number of hardware-triggered events (labelled ``total events''),
number of
  events with subtraction of 
  the expected number of 
accidental events and a correction for the overall dead time 
applied(labelled ``total
corrected events'')}
\label{accident}
\vspace{0.1cm}
\begin{center}
\begin{tabular}{lcccccc}
Crab  &  Total events & Tot. correct. evs \\
\hline
\hline
ON  & 79194   & 58107 \\
OFF & 86428   & 58550 \\
\hline
EXCESS & -7234 $\pm$ 407 &   -443 $\pm$ 341  \\
\hline
\end{tabular}
\end{center}

\begin{center}
\begin{tabular}{lcccccc}
3C454  &  Total events & Tot. correct. evs \\
\hline
\hline
ON  & 49141   & 49139 \\
OFF & 51982   & 49521 \\
\hline
EXCESS & -2841 $\pm$ 318 &   -382 $\pm$ 314  \\
\hline
\end{tabular}
\end{center}

\begin{center}
\begin{tabular}{lcccccc}
3EG+1835  &  Total events & Tot. correct. evs \\
\hline
\hline
ON  & 50264   & 53760 \\
OFF & 50914   & 54702 \\
\hline
EXCESS & -650 $\pm$ 318 & -942 $\pm$ 329  \\
\hline
\end{tabular}
\end{center}

\begin{center}
\begin{tabular}{lcccccc}
pseudo source  &  Total events & Tot. correct. evs \\
\hline
\hline
ON  & 28808   & 26010 \\
OFF & 31993   & 26549 \\
\hline
EXCESS & -3185 $\pm$ 246 & -539 $\pm$ 229  \\
\hline
\end{tabular}
\end{center}

\normalsize             
\end{table}
\\
In the alternative approach of applying a software threshold (
section \ref{softp})
accidental events are rejected due to their incorrect timing pattern.
Less than 0.6$\%$ of the accidental events pass the
analysis for
a very similar NSB to the one of Crab sample. In our
analysis, a higher NSB rejects more accidental events. As seen in the
column 6 of table \ref{crabresult} also here there is no significant
excess in the total rate.
This lack of an excess in the total rate seems to cast a doubt on the
reality
of the signal discussed in section \ref{results}.

\section{Conclusion}
\label{concl}
The results of the present paper do not
yet prove that the use
of an heliostat array in gamma-ray astronomy is a feasible alternative
to the use of dedicated Cherenkov telescopes. The properties
of experimentally detected showers---while
showing statistically significant deviations from Monte-Carlo
simulated proton showers---agree in some important derived
parameters to within typically 10$\%$.
Systematic errors were within the range expected.
Both the capital costs of the experiment and the running costs of remote 
night-time operation in a facility
used for solar-energy research during daytime 
are lower than of a comparable dedicated telescope.
\\
The principle drawbacks of this approach were found
to be the restricted field of view and the night-time weather conditions 
at the relatively low elevation of the heliostat field.
The field-of view restriction was shown to lead to a very
similar time structure of the shower front in proton and gamma induced
showers. Moreover it biases the direction reconstruction based in timing
towards to the pointing direction. 
Both effects together prevent an efficient
separation of proton and gamma induced showers. 
This makes a flux determination independent of total rates difficult
(though not impossible) and severely limits the 
sensitivity of the experiment.
It was found that the fraction of time (total duty cycle)
with weather and moon-light conditions sufficient
for the detection of gamma radiation was about 3-4$\%$ at the PSA, about
a factor 5 lower than at astronomical sites. 
Both drawbacks seem to be unavoidable also in the future, because an
enlargement of the field of view seems impossible within a budget
lower than for dedicated telescopes, and the site-selection criteria
for solar-energy generation facilities (e.g. low average wind speeds)
exclude astronomical sites in principle. The effect of different
night-sky backgrounds in the ON- and OFF-source region on the hardware
trigger threshold is still small for the relatively small event
numbers discussed in this paper and the observed maximal difference
of NSB intensity of 5$\%$.
However, this effect becomes a principal
difficulty for the determination of absolute fluxes in somewhat
larger samples.

{\bf Acknowledgments}
The GRAAL project is supported by funds from the DFG, CICYT and the 
EC-DGXII's 'Improving Human Potential programme
``Access to large scale facilities''.
We thank the PSA---in particular A.Valverde---for excellent working 
conditions at the CESA-1 heliostat field. We thank the
"Centro de Supercomputacion UCM" for support.
\vskip -2.0 cm
\pagebreak

\end{document}